\documentclass[12pt,preprint]{aastex}

\usepackage{graphics,epsfig}
\usepackage{natbib}

\def\spose#1{\hbox to 0pt{#1\hss}}
\def\lta{\mathrel{\spose{\lower 3pt\hbox{$\mathchar"218$}}
     \raise 2.0pt\hbox{$\mathchar"13C$}}}
\def\gta{\mathrel{\spose{\lower 3pt\hbox{$\mathchar"218$}}
     \raise 2.0pt\hbox{$\mathchar"13E$}}}

\def\figure#1#2 {\par{\narrower\noindent {\bf Fig. #1}
   \hskip 2mm #2\par}\bigskip\noindent}
\def\table#1#2 {\par{\narrower\noindent {\bf Tab. #1}
   \hskip 2mm #2\par}\bigskip\noindent}

\shorttitle{Habitable Trojan Planets}

\shortauthors{Eberle et al.}

\begin{document}


\title{Case Studies of Habitable Trojan Planets in the System of HD~23079}


\author{J. Eberle, M. Cuntz, B. Quarles, Z. E. Musielak}
\vspace{2.5cm}
\affil{Department of Physics, University of Texas at Arlington, Box 19059, \\
       Arlington, TX 76019, USA}
\vspace{0.5cm}
\email{wjeberle@uta.edu, cuntz@uta.edu, billyq@uta.edu, zmusielak@uta.edu}


\clearpage


\begin{abstract}
We investigate the possibility of habitable Trojan planets in the HD~23079
star-planet system.  This system consists of a solar-type star and a
Jupiter-type planet, which orbits the star near the outer edge of the
stellar habitable zone in an orbit of low eccentricity.  We find that
in agreement with previous studies Earth-mass habitable Trojan planets
are possible in this system, although the success of staying within the
zone of habitability is significantly affected by the orbital parameters
of the giant planet and by the initial condition of the theoretical
Earth-mass planet.  In one of our simulations, the Earth-mass planet
is captured by the giant planet and thus becomes a habitable moon.
\end{abstract}



\clearpage


\clearpage

\section{Introduction}

The existence of planets in orbit about solar-type stars is 
now a well-established observational result.  Obviously, the
ultimate quest of these studies is to discover Earth-like planets
located in the habitable zones (HZs) of their host stars.  So far,
a small number of super-Earth planets with masses of up to about
12 $M_\oplus$ \citep[e.g.,][]{udr07,vog10} have been found,
typically located around M-type dwarf stars.  Nevertheless,
the existence of Earth-mass planets, including those hosted
by solar-type stars, is strongly implied by various observational 
findings including the occurrence and mass distribution of
close-in super-Earths, Neptunes, and Jupiters \citep{how10}.
Measurements by the authors indicate an increasing planet
occurrence with decreasing planetary mass $M_p$ akin to $M_p^{-0.48}$,
implying that 23\% of stars harbor a close-in Earth-mass planet
(ranging from 0.5 to 2.0 $M_\oplus$); see also \cite{mar00} for
earlier results.  Very recent support for the existence
of Earth-type planets outside the Solar System is lent by the
discovery of Kepler-10b, Kepler's first rocky planet with an estimated
mass of 4.6 Earth masses \citep{bat11}.

Long-term orbital stability of Earth-like planets in stellar HZs is
necessary for the evolution of any form of life, particularly intelligent
life.  There is a large array of studies focusing on the orbital
stability of hypothetical Earth-mass planets in stellar HZs concerning
different types of host stars and star-planet configurations.  Examples
include studies by \cite{geh96}, \cite{jon01,jon05,jon06}, \cite{jon10},
\cite{nob02}, \cite{men03}, \cite{cun03}, \cite{blo03}, and \cite{asg04}.
Particular types of systems are those where a Jupiter-type planet
orbits a star in the stellar HZ, therefore jeopardizing the possibility
of habitable terrestrial planets in that system.  This is actually
the situation of HD~23079, the focus of the present paper.

Previously, \cite{nob02} investigated the orbital stability
of terrestrial planets inside the HZs of 47~UMa and HD~210277.
The center stars of these systems are very similar
to the Sun concerning mass, spectral type, and
effective temperature.  Orbital stability was attained for
the inner part of the HZ of 47~UMa; however, no orbital
stability was found for hypothetical Earth-mass planets in the HZ
of HD~210277.  In this case, a Jupiter-type planet crosses
the stellar HZ, thus effectively thwarting habitability for
this system.  Very recent examples were also given by \cite{yea11}
who studied the star-planet systems HD~20782 and HD~188015.
In both cases, the giant planet significantly interferes with
any Earth-mass planet in the stellar HZ assumed to have
formed there for the sake of study.  In all cases, the
Earth-mass planet was ejected from the stellar HZ in a
very short time.

However, if a giant planet is orbiting the star in the stellar
HZ, there is still the principal possibility of habitable
Trojan planets in those systems as pointed out by, e.g.,
\cite{dvo04} and \cite{sch07}.  A Trojan planet is one located
around one of the Lagrangian points L4 and L5 of the giant planet.
These points lie on the giant planet's orbit, ahead (L4) and
behind (L5) the planet, each forming an equilateral triangle
with the planet and its star.  Thus, Trojan planets are also
in a 1:1 resonance with the giant planet.  \cite{dvo04}
investigated the stability regions of hypothetical terrestrial
planets around L4 and L5 in specific systems,
including HD~23079, in the framework of the restricted
three body problem.  They obtained relationships between
the size of the stability regions and the orbital parameters
of the giant planets, particularly its eccentricity.  Studies
about Neptune Trojans were given by, e.g., \cite{dvo07}.

A study by \cite{sch07} identified several exoplanetary
systems that can harbour Trojan planets with stable orbits in the
stellar HZs.  Concerning HD~23079, this study concluded that
a Trojan planet will only spend 35\% of its time in the stellar HZ,
assumed to extend from 0.85 to 1.60~AU.  In our study, however,
we will consider a zone of habitability based on the generalized
estimate by \cite{und03}, implying habitability between
0.99 and 1.97~AU (see below).  This means that the habitable area,
which is the area of the stellar HZ annulus,
is increased by 58\% compared to that considered by \cite{sch07}.
In the present study, we will conclude that habitable Trojan
planets are indeed possible in the system of HD~23079, although
their existence is significantly affected by, e.g., the orbital
parameters of the giant planet.  Next we will describe our
theoretical approach.  We will discuss the adopted methods and
the system parameters for HD~23079.  Thereafter, we will describe
our results.  Finally, we will present our conclusions.

%%%%%%%%%%%%%%%%%%%%%%%%%%%%%%%%%%%%%%%%%%%%%%%%%%%%%%%%%%%%%%%%%%%%%%%%

\section{Theoretical Approach}

\subsection{Stellar and Planetary Parameters}

HD~23079 has been monitored as part of the Anglo-Australian Planet
Search (AAPS) program \citep{tin02} that is able to perform extrasolar
planet detection and measurements with a long-term, systematic radial
velocity precision of 3 m~s$^{-1}$ or better.  HD~23079 was identified
to host a Jupiter-type planet in a relatively large and nearly circular
orbit.  HD~23079 is an inactive main-sequence star; \cite{gra06}
classified it as F9.5~V (see Table~1; all parameters have their usual
meaning), an updated result compared to \cite{hou75} who found that
HD~23079 is intermediate between an F8 and G0 star.  Its stellar
spectral type corresponds to a mass of $M = 1.10 \pm 0.15$ $M_\odot$.
The stellar effective temperature and radius are given as
$T_{\rm eff} = 6030 \pm 52$ K and $R = 1.106 \pm 0.022$ $R_\odot$,
respectively \citep{rib03}.  Thus, HD~23079 is fairly similar to
the Sun, though slightly hotter and slightly more massive.
The detected planet (HD~23079~b) has a minimum mass of
$M_p \sin i = 2.45 \pm 0.21$ $M_J$.  Furthermore, it has a
semimajor axis of $a_p = 1.596 \pm 0.093$ AU and an eccentricity
of $e_p = 0.102 \pm 0.031$ \citep{but06}, corresponding to an
orbital period of $P = 730.6 \pm 5.7$ days.  The original results by
\cite{tin02} indicated very similar planetary parameters.

The orbital parameters of HD~23079~b are relatively similar to those
of Mars, implying that HD~23079~b is orbiting its host star in or
near the outskirts of the stellar HZ; see discussion
below.  The existence of HD~23079~b, a planet even more massive than
Jupiter, makes it difficult for a terrestrial planet to orbit HD~23079
at a similar distance without being heavily affected by the giant
planet; see results from previous case studies by \cite{nob02} and
\cite{yea11} who focused on the dynamics of HD~20782, HD~188015,
and HD~210277.  Concerning HD~23079, a previous investigation
pertaining to habitable terrestrial Trojan planets was given by
\cite{dvo04}.

%%%%%%%%%%%%%%%%%%%%%%%%%%%%%%%%%%%%%%%%%%%%%%%%%%%%%%%%%%%%%%%%%%%%%%%%

\subsection{Method of Integration}

For our simulations of the HD~23079 system, we consider both the
observed giant planet and a hypothetical terrestrial planet of
one Earth-mass, i.e., $3.005 \times 10^{-6}$~$M_\odot$, which allows
us to execute a grid of model simulations.  The method of integration
uses a fourth-order Runge-Kutta integration scheme \citep{gar00}.
The code has been extensively tested against known analytical solutions,
including the two-body and restricted three-body problem \citep[see][for
detailed results]{nob02,cun07,ebe08}.  In the framework of our simulations
that we limit to $10^6$~yrs, we apply a time-step of $10^{-4}$~yrs for
the integration scheme that is found to be fully appropriate.  In that
regard, we pursued test studies comparing the planetary orbits based on
three different integration time-steps, which are: $10^{-3}$, $10^{-4}$
and $10^{-5}$~yrs.  In particular, we evaluated $\Delta R_{ij}$, i.e.,
the magnitude of the difference between the position of the planet when
different step sizes of $10^{-i}$ and $10^{-j}$ were used.  We found
that there is no significant change in outcome between models with
time-steps of $10^{-4}$ and $10^{-5}$~yrs.

The initial conditions (i.e., starting velocities) for the orbits of the
Earth-mass planets were chosen such that the planet was assumed to
start at the midpoint of the stellar HZ (1.4779 AU) and to be in a circular
orbit about the star, although it is evident that it will be
significantly affected immediately by gravitational pull of the giant
planet, which will prevent the planet from continuing a circular motion.
For each set of models, defined by sets of values for the semimajor
axis $a_p$ and eccentricity $e_p$, given as 1.503, 1.596, 1.689~AU and
0.071, 0.102, 0.133, respectively, 8 different configurations are
considered.  They are defined by the 8 different starting (phase) angles
for the Earth-mass planet, which are varied in increments of 45$^\circ$
noting that 0$^\circ$ corresponds to the 3~o'clock position.  Moreover,
the starting position of the Jupiter-type planet (HD~23079~b), for which
we assume its minimum mass value of 2.45 $M_J$, was varied between its
periastron and its apastron position.  Therefore, a total of 144 initial
configurations has been considered.  Note that the Jupiter-type planet
was always started at the 3~o'clock position, which after adjusting the
orbital layout of the giant planet always coincided with its periastron
(see Fig.~1) or apastron position depending on the type of model.  

%%%%%%%%%%%%%%%%%%%%%%%%%%%%%%%%%%%%%%%%%%%%%%%%%%%%%%%%%%%%%%%%%%%%%%%%

\subsection{Stellar Habitable Zone}

The extent of the HZ of HD~23079 has been calculated following
the formalism by \cite{und03} based on previous work by
\cite{kas93}.  \cite{und03} supplied a polynomial fit
depending on the stellar luminosity and the stellar effective
temperature that allows to calculate the extent of the
conservative and the generalized HZ.  Noting that HD~23079
is more luminous than the Sun, it is expected that its HZ
is more extended than the solar HZ, for which the inner and
outer limit of the generalized HZ were given as 0.84 and 1.67~AU,
respectively \citep{kas93}.
The generalized HZ is defined as bordered by the runaway
greenhouse effect (inner limit), where water vapour enhances
the greenhous effect thus leading to runaway surface warming,
and by the maximum greenhouse effect (outer limit), where a
surface temperature of 273~K can still be maintained by a
cloud-free CO$_2$ atmosphere.  The inner limit of the
conservative HZ is defined by the onset of water loss, i.e.,
the atmosphere is warm enough to allow for a wet stratosphere
from where water is gradually lost by photodissociation and
subsequent hydrogen loss to space.  Furthermore, the outer
limit of the conservative HZ is defined by the first CO$_2$
condensation attained by the onset of formation of CO$_2$
clouds at a temperature of 273~K.

For HD~23079, the limits of the conservative HZ are given
as 1.1378 and 1.6362~AU, whereas the limits of generalized HZ
are given as 0.9896 and 1.9662~AU (see Fig. 1).  The limits
of the generalized HZ are those employed in our numerical
planetary studies\footnote{The physical limits
of habitability are much less stringent than implied by the
numerical precision of these values; nevertheless, these values
were used for checking if the Earth-mass planet has left
the stellar HZ.}.  The underlying definition of habitability
is based on the assumption that liquid surface water is a
prerequisite for life, a key concept that is also the basis
of ongoing and future searches for extrasolar habitable planets
\citep[e.g.,][]{cat06,coc09}.  The numerical evaluation
of these limits is based on an Earth-type planet with a
CO$_2$/H$_2$O/N$_2$ atmosphere.  Specifically, the inner limit
of habitability is set by the loss of water from
the upper planetary atmosphere through photodissociation and
subsequent escape of hydrogen to space associated with a
run-away greenhouse effect.  The outer limit of habitability
is given by the maximum greenhouse effect \citep{kas93,und03},
by which a surface temperature of 273~K can be maintained by a
cloud-free CO$_2$ atmosphere.

We point out that concerning the outer edge of habitability,
even less conservative limits have been proposed in the meantime
\citep[e.g.,][]{for97,mis00}.  They are based on the assumption of
relatively thick planetary CO$_2$ atmospheres and invoke strong
backwarming that may further be enhanced by the presence of CO$_2$
crystals and clouds.  However, as these limits, which can be as
large as 2.4~AU in case of the Sun, depend on distinct properties
of the planetary atmosphere, they are not relevant for our study.
Nevertheless, we convey this type of limit for the sake of curiosity
(see Fig. 1), noting that it has properly been adjusted to 2.75~AU
in consideration of the radiative conditions of the planetary host star,
HD~23079.  Moreover, the significance of this extreme limit
has recently been challenged based on detailed radiative transfer
simulations \citep{hal09}.

%%%%%%%%%%%%%%%%%%%%%%%%%%%%%%%%%%%%%%%%%%%%%%%%%%%%%%%%%%%%%%%%%%%%%%%%

\section{Results and Discussion}

\subsection{Case Studies of Habitable Trojan Planets}

Table~2 and Table~3 summarize the time the Earth-mass planet
remains within the stellar HZ, i.e., before exiting the stellar HZ
or being permanently ejected from the system.  Of the 144 total
considered initial configurations 13 survived at least 1 million years,
93 crossed the upper limit of the HZ, 28 crossed the lower limit of
the HZ, and 10 collided or may have had a very close approach with
the giant planet.  Some of those who crossed the HZ at the lower or
upper limit as first exit from the HZ may have had a very close
approach with the giant planet, resulting in destruction while
entering the Roche limit \citep{wil03} or in a collision with the
giant planet, at a later time.  Of the 13 survivors,
12 are Trojan types, that is they exist in stable orbits around
the equilateral equilibrium positions much like that
demonstrated in \cite{dvo04}.

In the cases where the giant planet is initially in the periastron
position, only models with the smallest considered semimajor axis
and eccentricity combination, which are $a_p = 1.503$~AU and
$e_p = 0.071$, result in habitable Trojan planets (see Fig.~2).
In this case, four different starting positions (phase angles)
appear to be consistent with long-term stability (see Table~2).
It is clear that the Earth-mass planet is safely inside of the
stellar HZ but it is a snug fit.
For the next larger eccentricity considered, which is 0.102,
there are various cases where the Earth-mass planet stays within
the HZ for some hundreds thousand years before finally crossing
the upper limit of the HZ.  When the eccentricity of the giant planet
is increased to 0.133, the Earth-mass planet remains within the HZ
at best for only a few hundred years.

The situation is, in principle, similar for the cases where the
giant planet is initially placed at the apastron position.
In this case, for $a_p = 1.503$~AU, the Earth-mass planet remained
in the stellar HZ for at least a million years for two eccentricity
simulations, which are $e_p = 0.071$ and 0.102 (see Figs.~3 and 4,
respectively).  Comparing Fig.~4 to Fig.~3, it is clear that the
Earth-mass planet moves in a wider area and approaches the edges
of the HZ for the larger eccentricity, thus illustrating how the
planet remained in the HZ for such a long time before exiting in
the periastron case with the same parameters.  In some of those
latter cases, we found that the planet was outside the HZ for
a brief time (i.e., considerably less than a planetary orbit),
but most likely without losing its habitability. This conclusion
is motivated by the previous study of \cite{wil02} that showed
that brief excursions from the HZ are insufficient to nullify
planetary habitability because the latter is expected to mainly
depend on the average stellar flux received over an entire orbit,
rather than the length of the time spent within the HZ.

Note that Figs.~2 and 3 display models of orbital stability for
the Earth-mass planet in a synodic (rotating) coordinate system.
Thus, the ``banana-shaped" areas correspond to the domains about
L4 or L5, where stability for the Earth-mass planet is encountered.
The thin line at the 3~o'clock position corresponds to the motion
of the giant planet due to its slightly elliptical orbit.
Clearly, only Earth-mass planets placed at phase angles of
45$^\circ$, 90$^\circ$, 270$^\circ$, and 315$^\circ$ have a
reasonable chance to develop into Trojan planets, whereas for
other starting angles ejections from the HZ, and usually also
from the star-planet system, will occur due to gravitational
interaction with the giant planet.  If the Earth-mass planet
was initially placed at an angle of 60$^\circ$ or 300$^\circ$,
it can be expected that it will continue to remain a Trojan planet.

For the sake of curiosity, we also evaluated various cases where the
Earth-mass planet never had a chance of becoming habitable.  Hence, we
chose five cases of different semimajor axes and eccentricities for the
giant planet.  In all cases the giant planet started at the periastron
position and the initial phase angle of the Earth-mass planet was chosen
as 180$^\circ$.  The simulations are depicted in Fig.~5.
In Fig.~5a, with $a_p = 1.503$~AU and $e_p = 0.102$, the system
experiences a relatively long period during which the Earth-mass planet first
exits the HZ at 48.4 yrs.  This event is preceded by a close approach
with the giant planet.  The simulation is terminated at 195.9 yrs due to an
expected collision with the giant planet.

Figures 5b to 5d are all based on $a_p = 1.596$~AU, but the depicted
simulations assume different eccentricities for the giant planet, which are
$e_p = 0.071$, 0.102, and 0.133, respectively.  In Fig.~5b, the system experiences
a short period during which the Earth-mass planet first exits the HZ at 7.93 yrs.
This event is again preceded by a close approach with the giant planet.  The
simulation is terminated at 9112 yrs due to the expected collision with
the giant planet.  In Fig.~5c, the system experiences
a short period during which the Earth-mass planet first exits the HZ at 7.80 yrs.
This event is preceded by a close approach with the giant planet.  The
simulation is terminated at 13400 yrs considering that the Earth-mass
planet is ejected from the system.  Habitability is ultimately prevented
as the Earth-mass planet becomes ``free-floating".  Free-floating planets
have previously been observed in case of the Trapezium cluster \citep{luc00};
note that planetary ejections due to orbital instabilities are an important
candidate process for this finding.

In Fig.~5d, with $a_p = 1.596$~AU and $e_p = 0.133$, this system experiences
a short period during which the Earth-mass planet first exits the HZ at 7.82 yrs.
It is reentering and exiting the HZ several times.  However, the simulation
is terminated at 54.60 yrs due to an expected collision with the giant planet.
In case of Fig.~5e, with $a_p = 1.689$~AU and $e_p = 0.102$, the system again
experiences a short period during which the Earth-mass planet first exits
the HZ at 4.78 yrs.  This event is preceded by a close approach with
the giant planet.  Eventually, the planet also becomes free-floating;
the simulation is terminated at about $7 \times 10^4$ yrs.
Figure 5a and 5d show oscillatory behaviours regarding the orbital motion
of the Earth-mass planet.  Noting that the Earth-mass planet starts
at a phase angle of 180$^\circ$, it initially orbits the star.  However,
when it approaches the giant planet, its orbit is being perturbed causing
the loops.  Thus, the Earth-mass planet exits and re-enters the HZ multiple
times until the end of the simulation.

\subsection{On the Possibility of Habitable Moons}

Our set of model simulations reveal a considerable variety in the
dynamics of the Earth-mass planet.  The most surprising case is
the following:  For $a_p = 1.596$~AU and $e_p = 0.133$ (see
Fig.~6) with the Jupiter-type planet initially placed at periastron
position and the Earth-mass planet placed at $0^\circ$, it was
found that the latter never crosses the inner or outer limit of
the stellar HZ during the simulation time of 10$^6$ years.  However, it
is found to orbit the giant planet in a retrograde orbit (relative
to the orbital motion of the giant planet about the star).
In this case, the Earth-mass planet is captured by the giant planet
and becomes a habitable moon, which occurs almost immediately after
the start of the simulation.

The analysis of its orbital data shows that the moon's semimajor
axis concerning its motion about the giant planet is $a_{\rm moon}
\simeq 0.051$~AU.  Its eccentricity is $e_{\rm moon} \simeq 0.8$
entailing a perigee and apogee of 0.0034 and 0.098~AU, respectively.
Thus, with a uniform data sampling rate, the moon is most likely to
be recorded at or near apogee.  From Fig.~7 it is evident that there
is also a precession of the perigee in a retrograde sense with a
period of approximately 30 years.  Figure 8 shows two histograms
regarding the time-dependent distance of the moon from the
giant planet, which reconfirms the moon's highly eccentric orbit.
The existence of a habitable moon in the HD~23079 system is also
consistent with the criterion of Hill stability as pointed out by,
e.g., \cite{don10}.  This study explores dynamic Hill stability
for a large variety of three-body systems considering moon/planet
mass ratios of 0.1, 0.01 and 0.001.

There is a persistent interest in the study of habitable
moons with respect to extrasolar giant planets orbiting host stars in
the stellar HZs.  Previous studies of habitable moons in systems akin
to HD~23079 have been given by \cite{wil97}, \cite{bar02}, and others.
The study by \cite{wil97} did not include HD~23079b as this
star-planet system was unknown at the time when this study was pursued.
However, by targeting the companions of 16~Cyg~B and 47~UMa,
\cite{wil97} investigated appropriate orbital parameters of
possible moons, and pointed out that the moons need to be large enough
(i.e., $> 0.12~M_\oplus$) to retain a substantial and long-lived
atmosphere, and furthermore would need to possess a significant
magnetic field to prevent its atmosphere from being sputtered away by
the ongoing bombardment of energetic ions from the planet's
magnetosphere, if existing.  Another study of possible moons, which
is fully applicable to the HD~23079 star-planet system, has been given
by \cite{bar02}.  They concluded that Earth-like moons of a Jovian planet
like HD~23079b would be able to exist for at least 5~Gyr considering
that the stellar mass of HD~23079 exceeds 0.15~$M_\odot$.

%%%%%%%%%%%%%%%%%%%%%%%%%%%%%%%%%%%%%%%%%%%%%%%%%%%%%%%%%%%%%%%%%%%%%%%%

\section{Conclusions}

The aim of our study was to add to the investigation of habitable Trojan
planets in the HD~23079 star-planet system.   This system consists of
a main-sequence star slightly hotter than the Sun.  Additionally,
it contains a Jupiter-type planet with a minimum mass of 2.45~$M_J$
that is orbiting the star in a slightly elliptical orbit that is
positioned within the stellar HZ.  The main goal of our study was
to explore if Earth-mass habitable Trojan planets can exist in this
system.

As the centerpiece of our study, we calculated a total of 144 orbital
stability simulations for the Earth-mass planet by choosing different
starting positions (phase angle) as well as placing the Jupiter-type planet
either at periastron or apastron position.  The attainment of habitability
solutions was found to critically depend on various parameters, which
include the orbital parameters of the giant planet (semi-major axis,
eccentricity) and the initial condition (phase angle) of the
theoretical Earth-mass planet.  We encoutered a variety of
different outcomes, which include (1) ejection of the Earth-mass planet
from the system, (2) engulfment of the planet by the star (or possible
destruction in accord with the Roche limit criterion), (3) capture of
the planet, thus becoming an habitable moon, or (4) remaining within the
stellar HZ.  The latter case was only attained in models where the orbit of the
giant planet had a relatively low eccentricity (but still within its
observationally given uncertainty), which however may be partially
due to the implemented choice of planetary starting positions.
Concerning the latter case, there were also cases (not shown in detail)
where the planet took short-term excursions from the HZ (i.e., considerably
less than the orbital period of HD~23079b, which is about 730~d), which
should be insufficient to nullify its habitability because the latter is
expected to mainly depend on the average stellar flux received over an
entire orbit, rather than the length of the time spent within the HZ
\citep{wil02}, although the ultimate effect of temporarily leaving the zone of
habitability will still partially depend on the atmospheric
thickness, structure and composition \citep[e.g.,][]{dre10}.

Moreover, we note that our study is supplementing previous work
by \cite{sch07} who concluded that a Trojan planet in the HD~23079
star-planet system will only spend 35\% of its time in the stellar HZ.
However, this estimation was based on a considerably narrower zone
of habitability than used in the present study.  Another, albeit minor,
difference is that \cite{sch07} used slightly different orbital parameters
for HD~23079b than in the current study.  In conclusion, it can be argued
that the system of HD~23079 is very well suited for the existence of
habitable Earth-type Trojan planets, and thus deserves serious
consideration in ongoing and future planetary search missions.

%%%%%%%%%%%%%%%%%%%%%%%%%%%%%%%%%%%%%%%%%%%%%%%%%%%%%%%%%%%%%%%%%%%%%%%%
%%%%%%%%%%%%%%%%%%%%%%%%%%%%%%%%%%%%%%%%%%%%%%%%%%%%%%%%%%%%%%%%%%%%%%%%

\bigskip
\bigskip
\bigskip

\centerline{\bf Acknowledgements}

This work has been supported by the U.S. Department of Education under GAANN
Grant No. P200A090284 (J.~E. and B.~Q.), the SETI institute (M.~C.) and
the Alexander von Humboldt Foundation (Z.~E.~M.).




%%clearpage to be removed

\clearpage

%%%%%%%%%%%%%%%%%%%%%%%%%%%%%%%%%%%%%%%%%%%%%%%%%%%%%%%%%%%%%%%%%

%%% *** Fig.1
%%%%%%%%%%%%%%%%%%%%%%%%%%%%%%%%%%%%%%%%%%%%%%%%%%%%%%%%%%%%%%%%%
\begin{figure*}
\centering
\epsfig{file=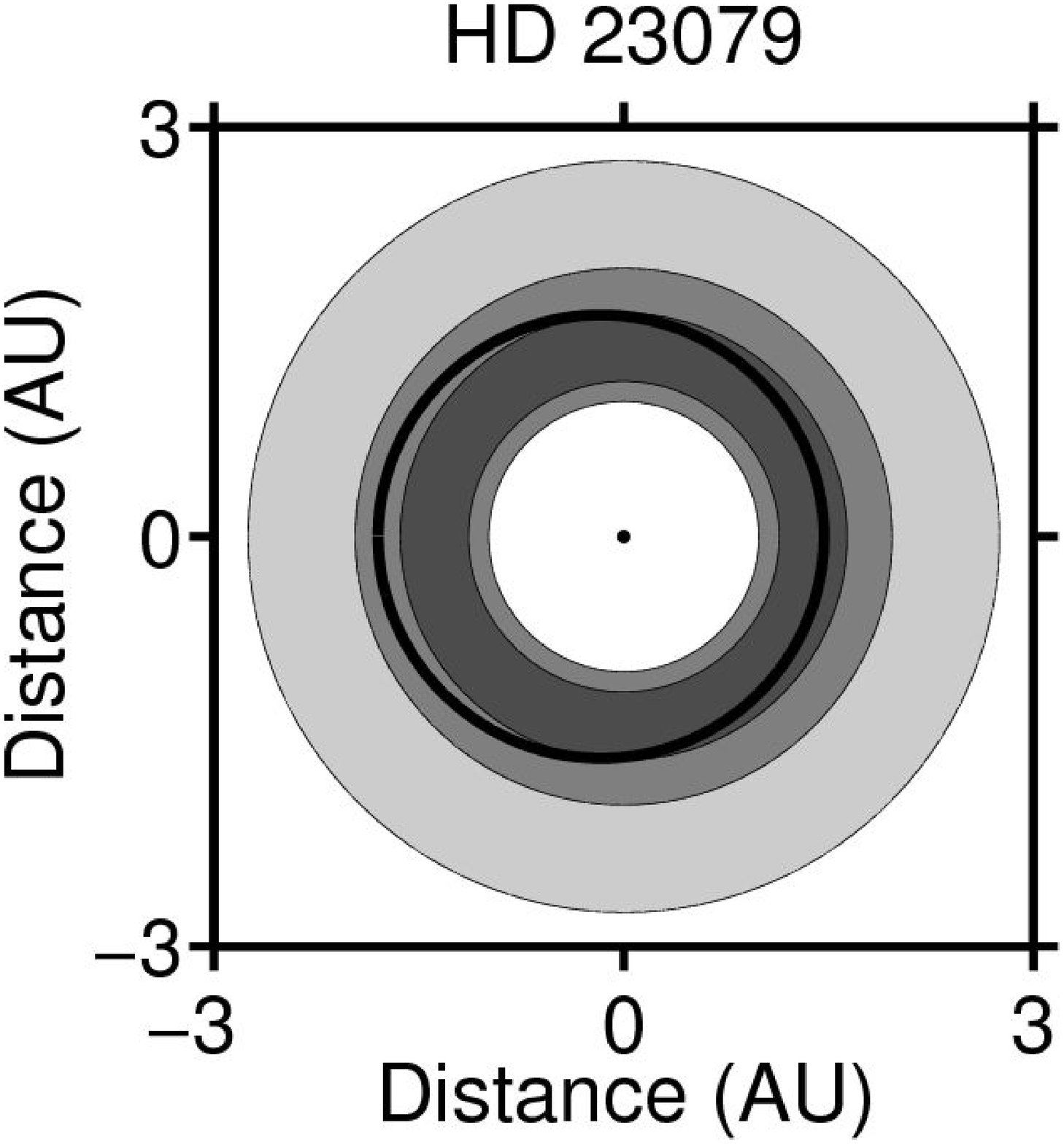,width=0.60\linewidth}
\caption{
Extent of the HZ for HD~23079, defined by its conservative limits
(dark grey) and generalized limits (medium grey).  In addition, we
depict the outer limit of an extreme version of the generalized HZ
(light grey) following the work by \cite{mis00}, although this limit
may be unrealistic based on subsequent studies.  The orbit of HD~23079~b,
a Jupiter-type giant planet, is depicted by a thick solid line.
}
\end{figure*}

%%%%%%%%%%%%%%%%%%%%%%%%%%%%%%%%%%%%%%%%%%%%%%%%%%%%%%%%%%%%%%%%%

\clearpage

%%% *** Fig.2
%%%%%%%%%%%%%%%%%%%%%%%%%%%%%%%%%%%%%%%%%%%%%%%%%%%%%%%%%%%%%%%%%
\begin{figure*}
\centering
\begin{tabular}{cc}
\epsfig{file=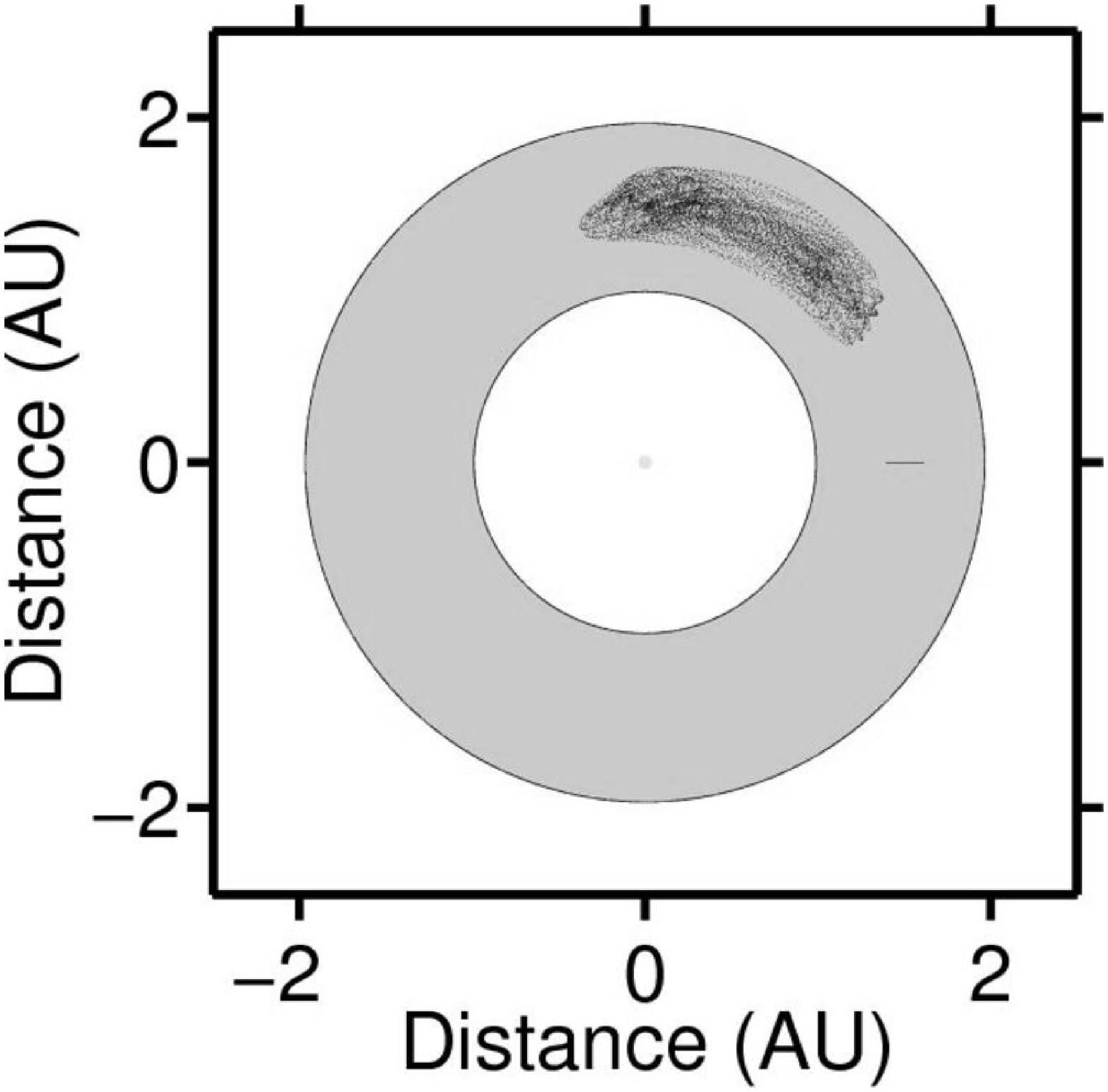,width=0.40\linewidth} &
\epsfig{file=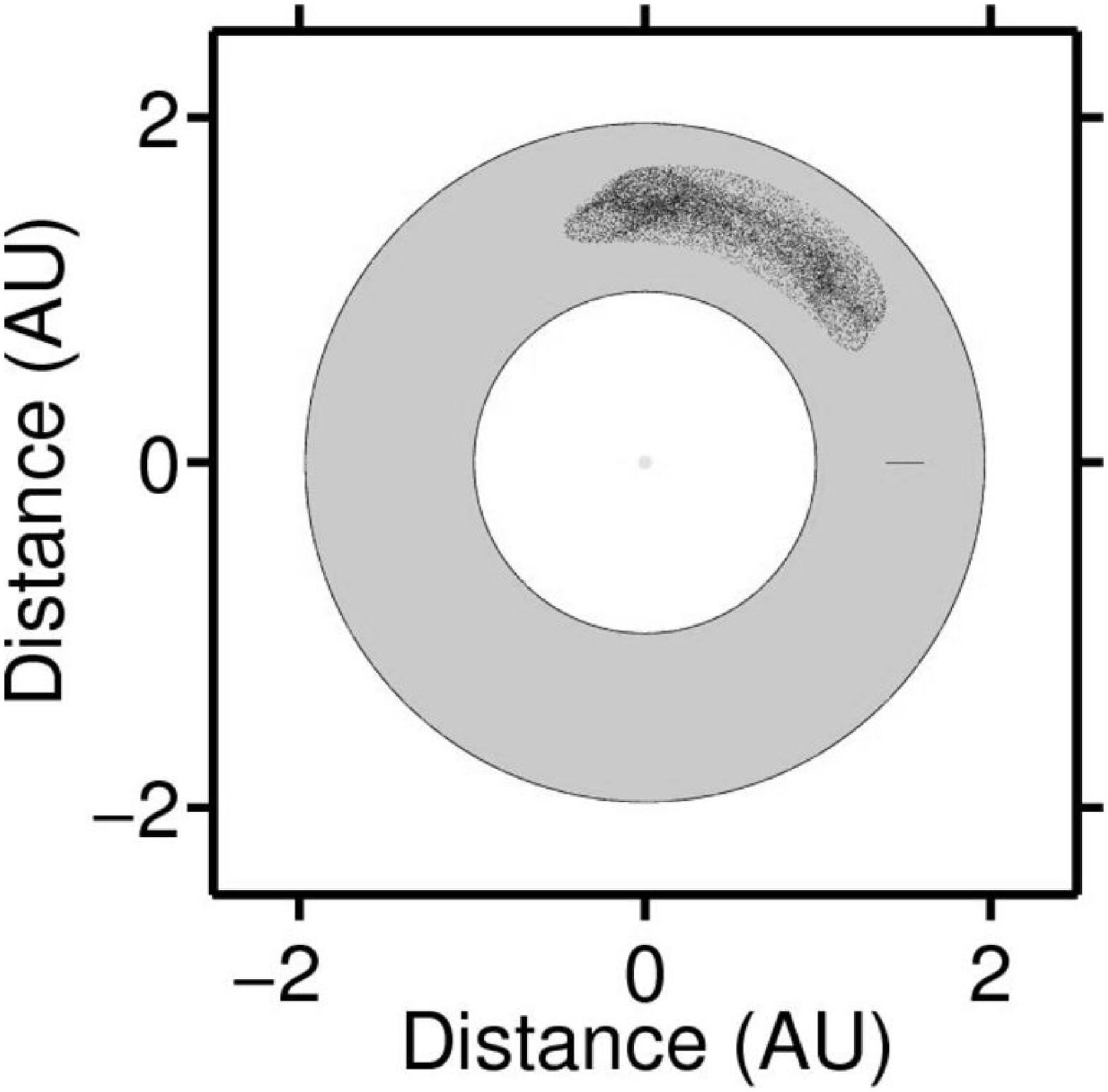,width=0.40\linewidth} \\
\epsfig{file=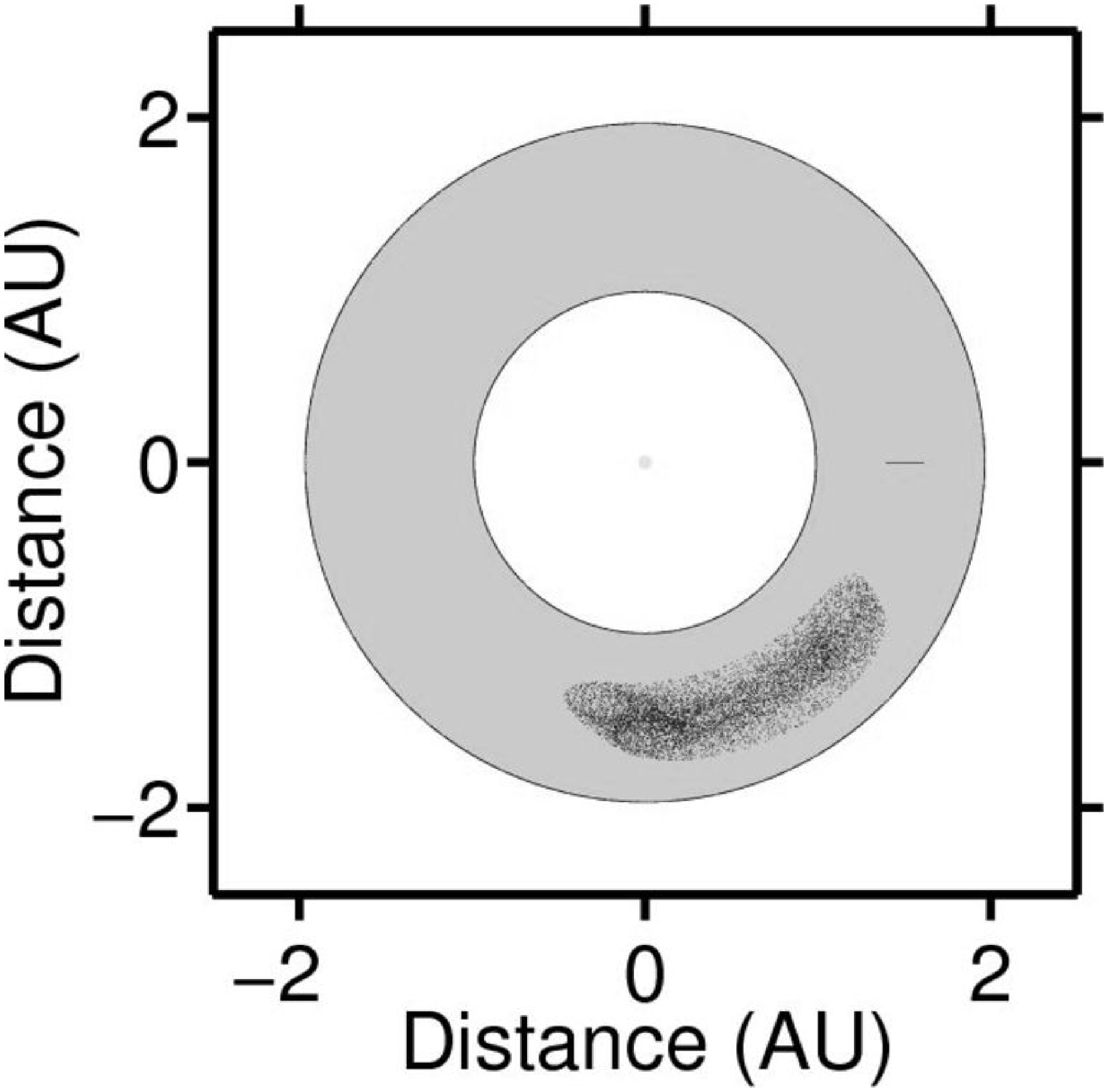,width=0.40\linewidth} &
\epsfig{file=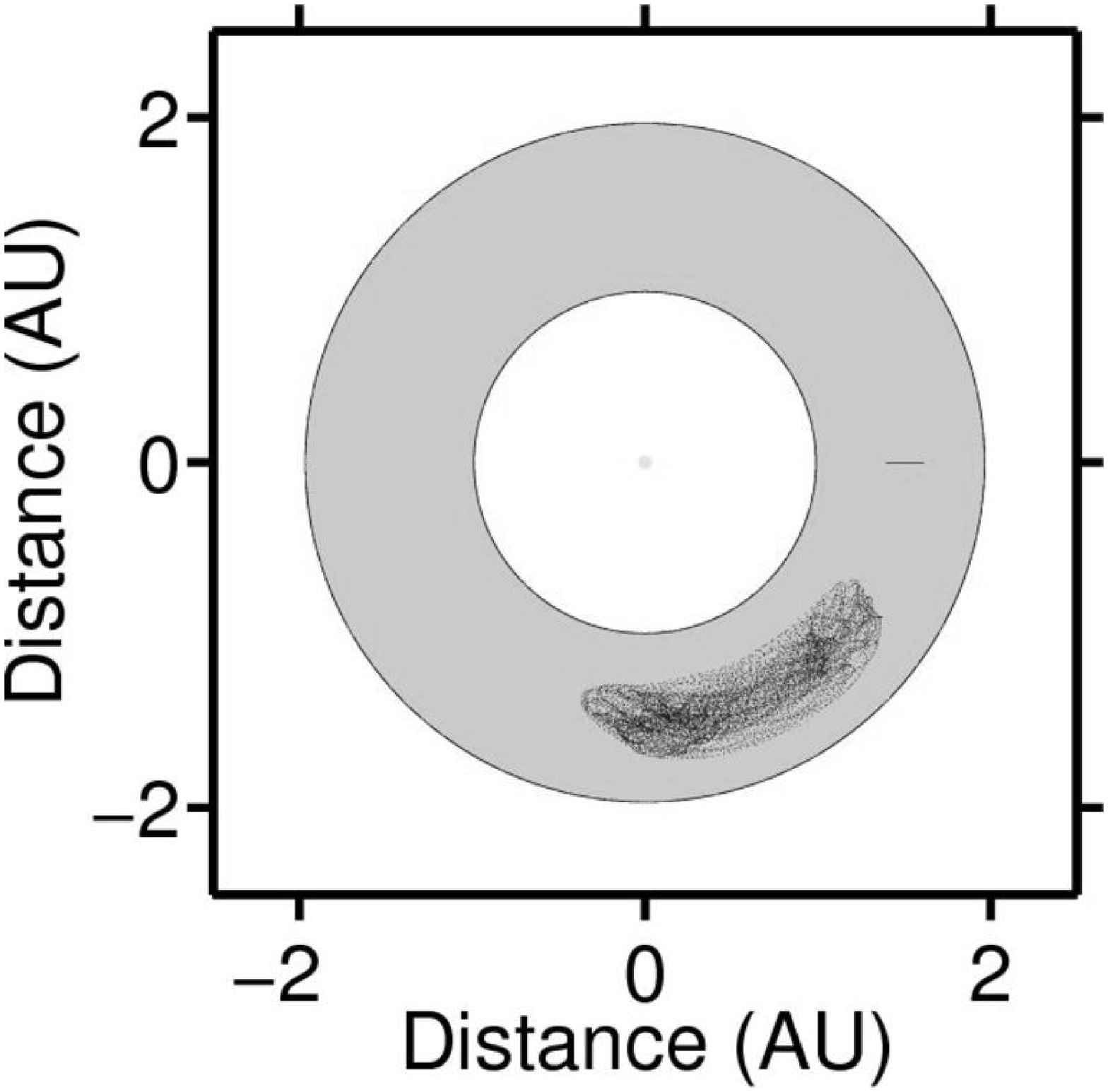,width=0.40\linewidth}
\end{tabular}
\caption{
Orbital stability simulations with HD~23079~b initially placed at periastron
with $a_p$ = 1.503~AU, $e_p$ = 0.071 and the Earth-mass planets placed at
four different starting angles, which are:
45$^\circ$ (top left), 90$^\circ$ (top right), 270$^\circ$ (bottom left),
and 315$^\circ$ (bottom right).  Using a rotating coordinate system,
HD~23079~b moves along the thin line.  The Earth-mass Trojan planets,
which give rise to the ``banana-shaped" area at L4 or L5, remain within
the HZ for at least 10$^6$ years.
}
\end{figure*}

%%%%%%%%%%%%%%%%%%%%%%%%%%%%%%%%%%%%%%%%%%%%%%%%%%%%%%%%%%%%%%%%%

\clearpage

%%% *** Fig.3
%%%%%%%%%%%%%%%%%%%%%%%%%%%%%%%%%%%%%%%%%%%%%%%%%%%%%%%%%%%%%%%%%
\begin{figure*}
\centering
\begin{tabular}{cc}
\epsfig{file=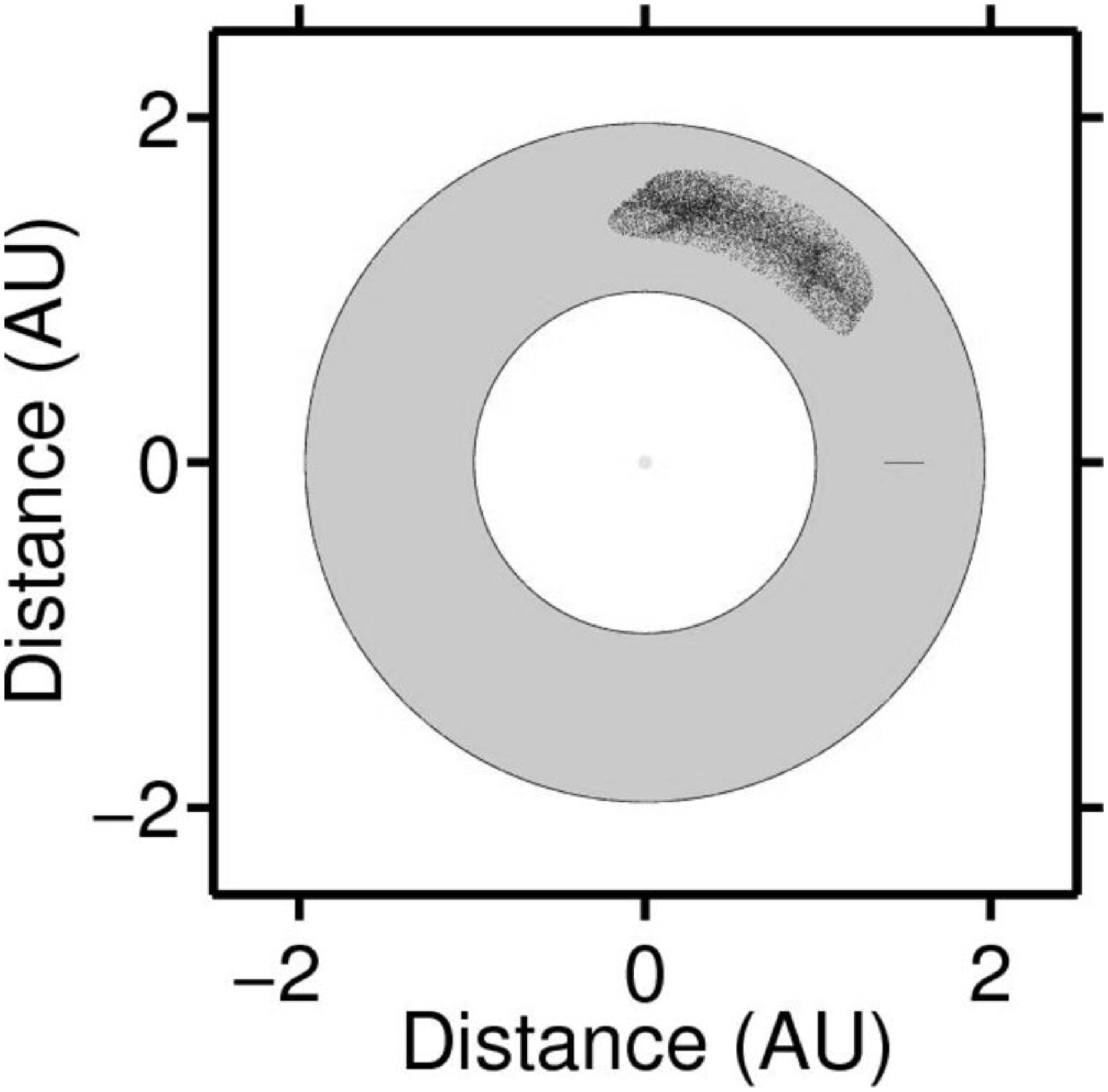,width=0.40\linewidth} &
\epsfig{file=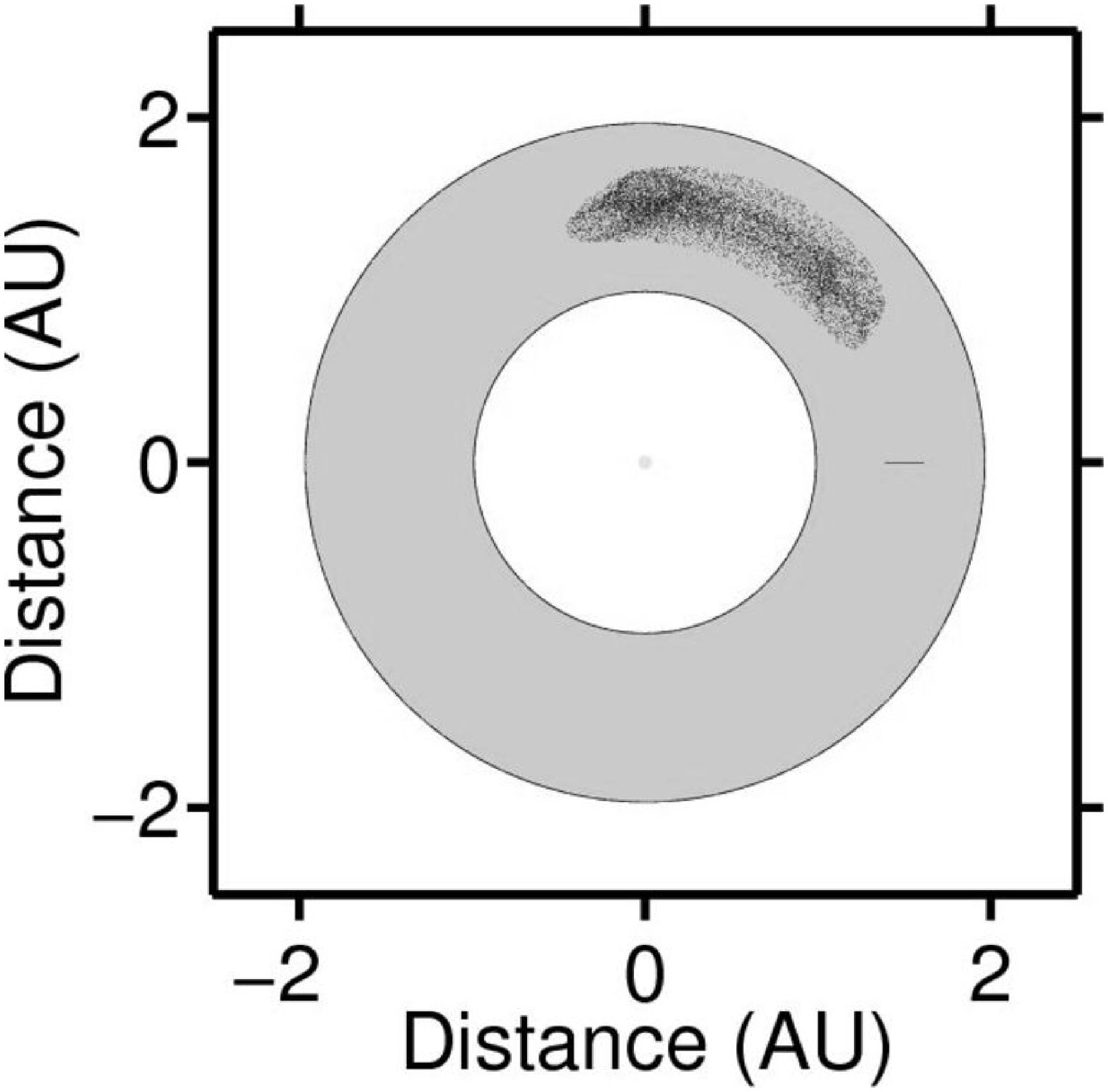,width=0.40\linewidth} \\
\epsfig{file=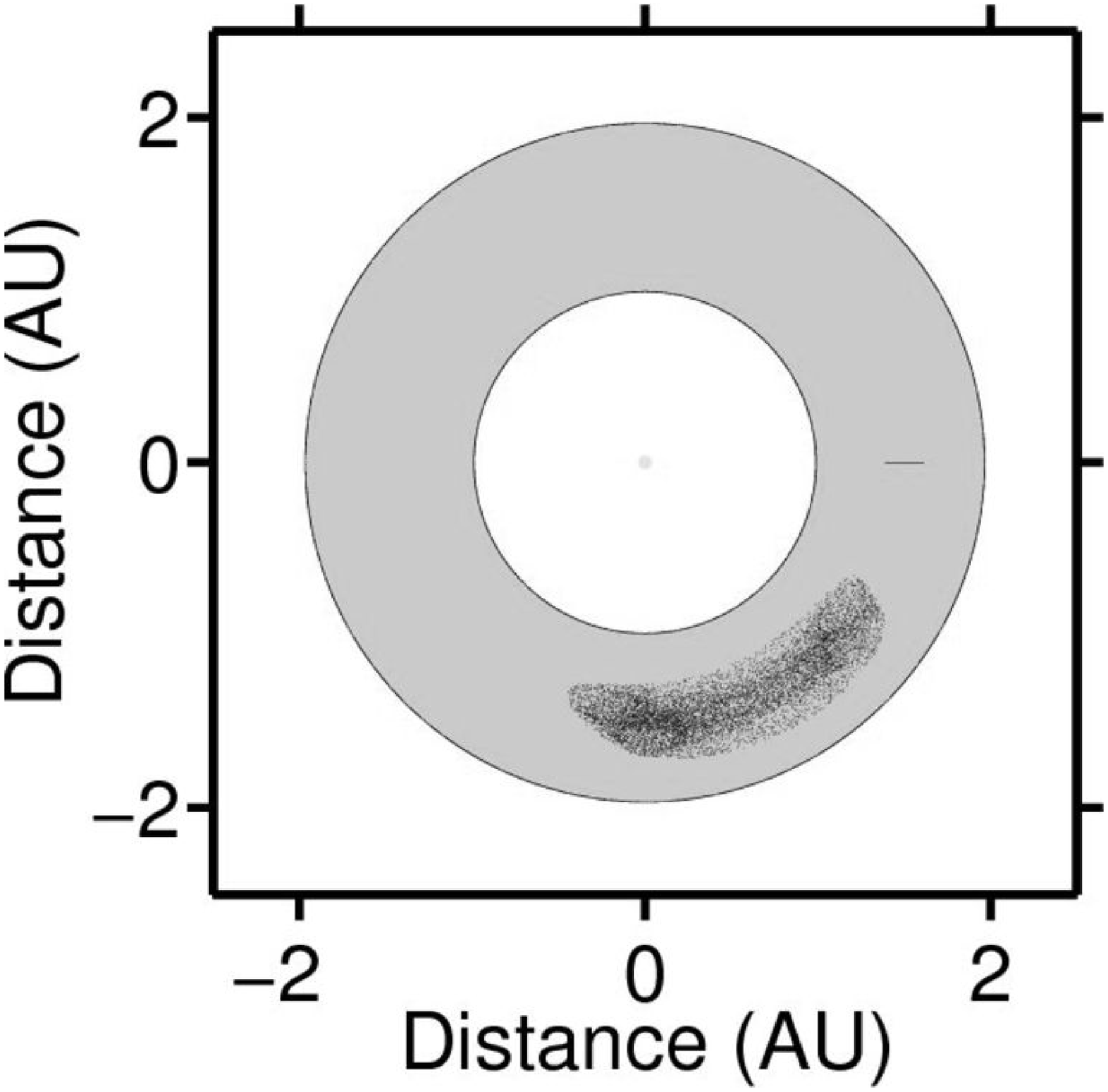,width=0.40\linewidth} &
\epsfig{file=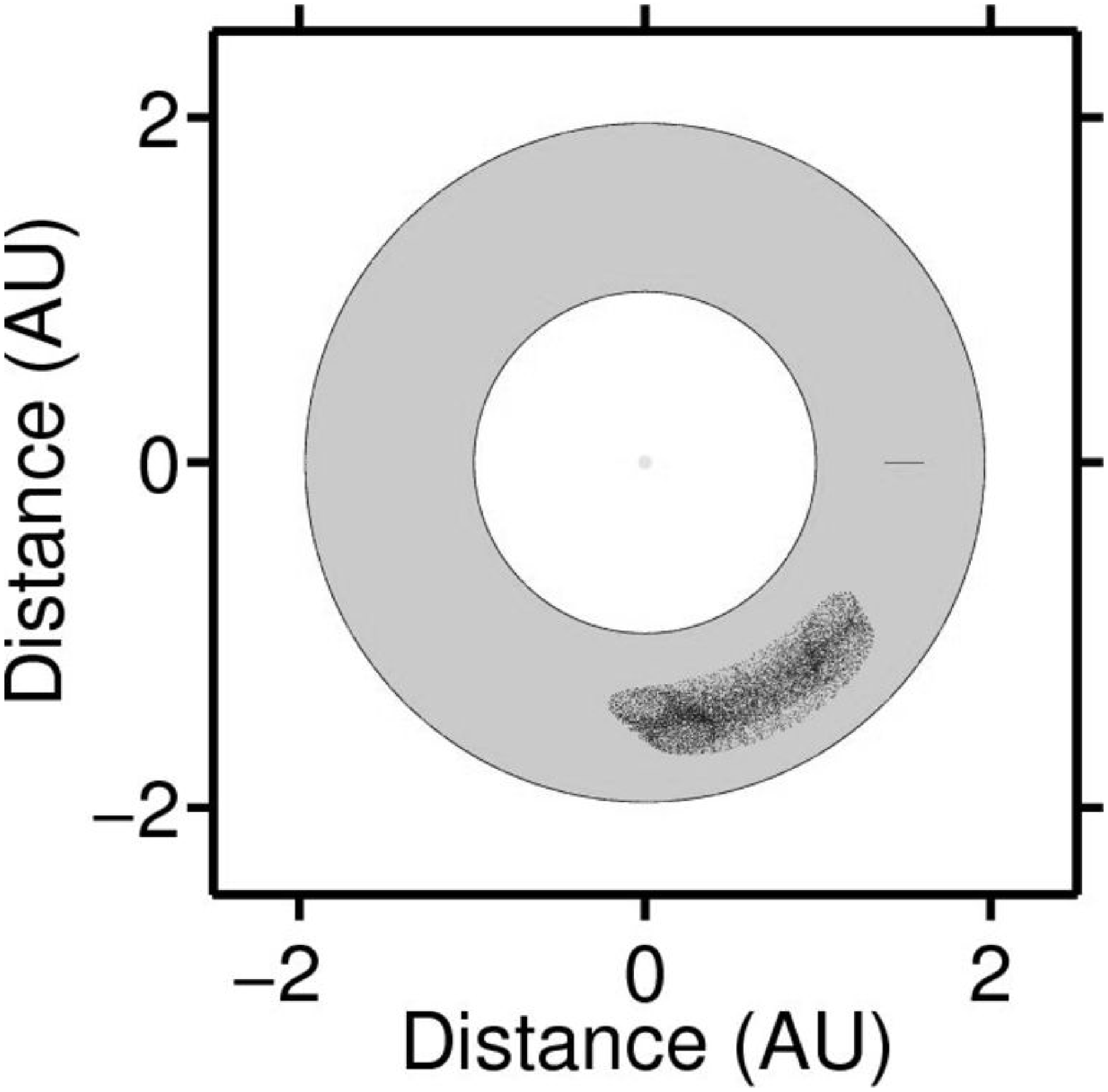,width=0.40\linewidth}
\end{tabular}
\caption{
Orbital stability simulations with HD~23079~b initially placed at apastron
with $a_p$ = 1.503~AU, $e_p$ = 0.071 and the Earth-mass planets placed at
four different starting angles, which are:
45$^\circ$ (top left), 90$^\circ$ (top right), 270$^\circ$ (bottom left),
and 315$^\circ$ (bottom right).  Using a rotating coordinate system,
HD~23079~b moves along the thin line.  The Earth-mass Trojan planets,
which give rise to the ``banana-shaped" area at L4 or L5, remain within
the HZ for at least 10$^6$ years.
}
\end{figure*}

%%%%%%%%%%%%%%%%%%%%%%%%%%%%%%%%%%%%%%%%%%%%%%%%%%%%%%%%%%%%%%%%%

\clearpage

%%% *** Fig.4
%%%%%%%%%%%%%%%%%%%%%%%%%%%%%%%%%%%%%%%%%%%%%%%%%%%%%%%%%%%%%%%%%
\begin{figure*}
\centering
\begin{tabular}{cc}
\epsfig{file=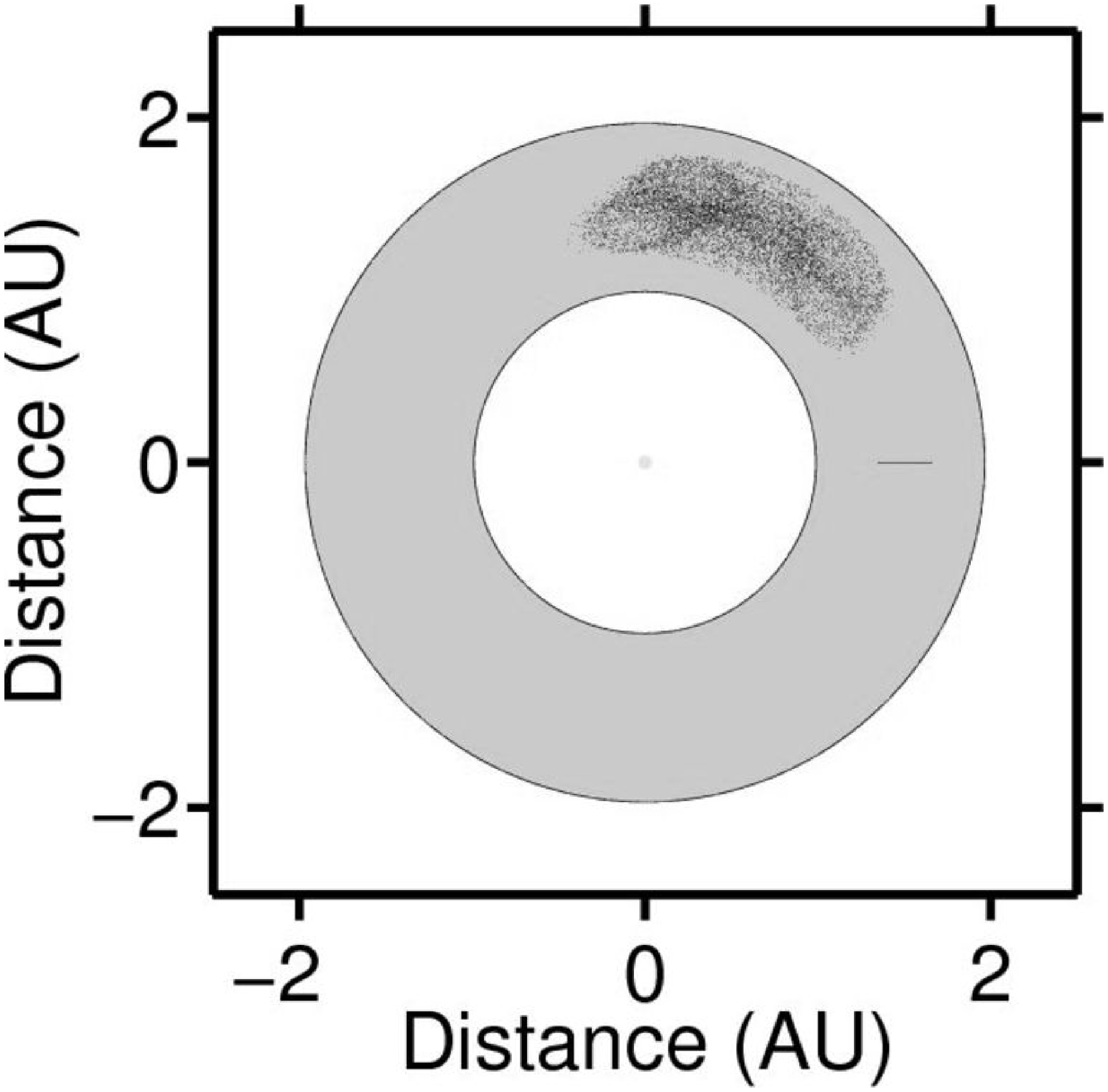,width=0.40\linewidth} &
\epsfig{file=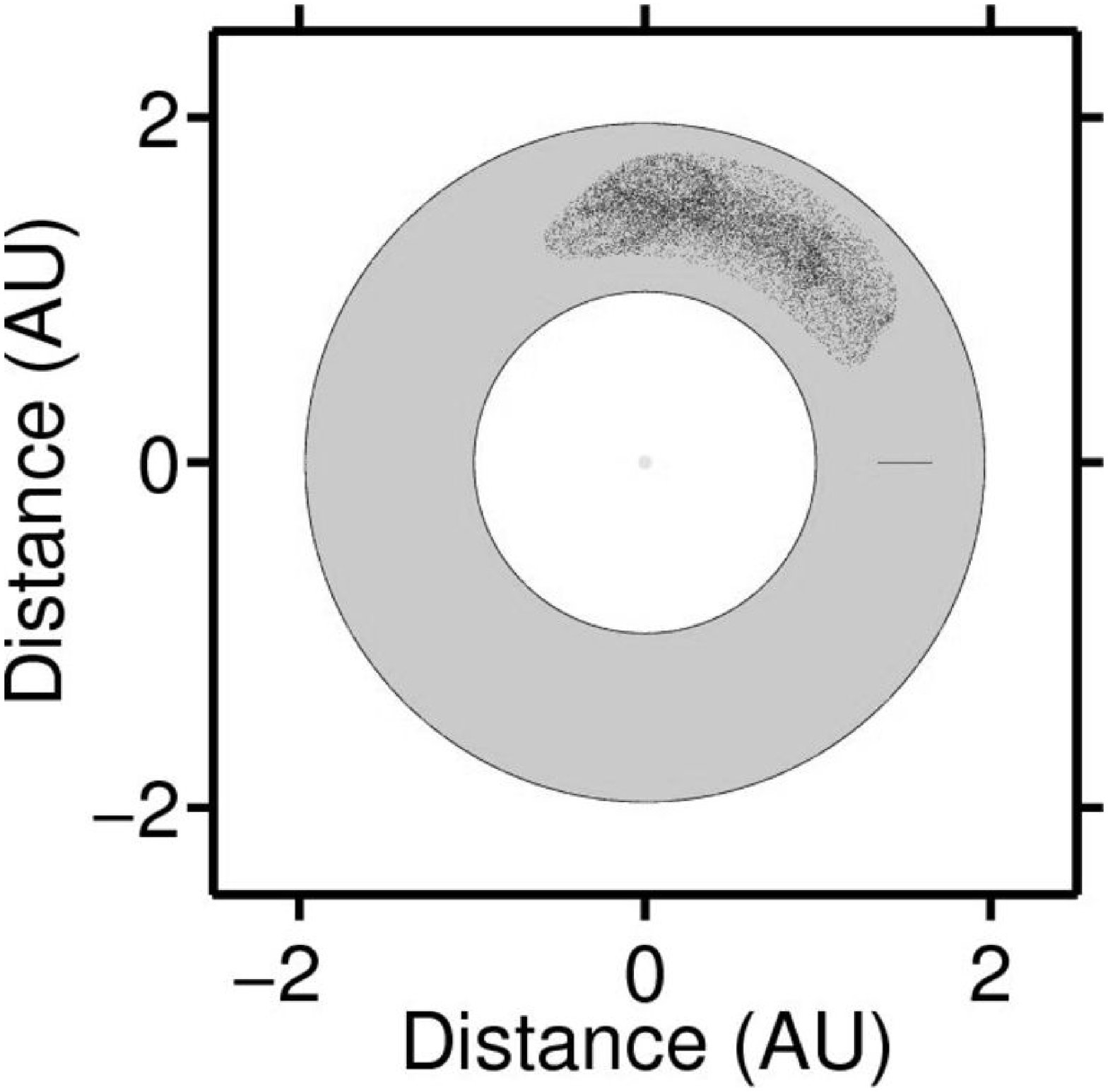,width=0.40\linewidth} \\
\epsfig{file=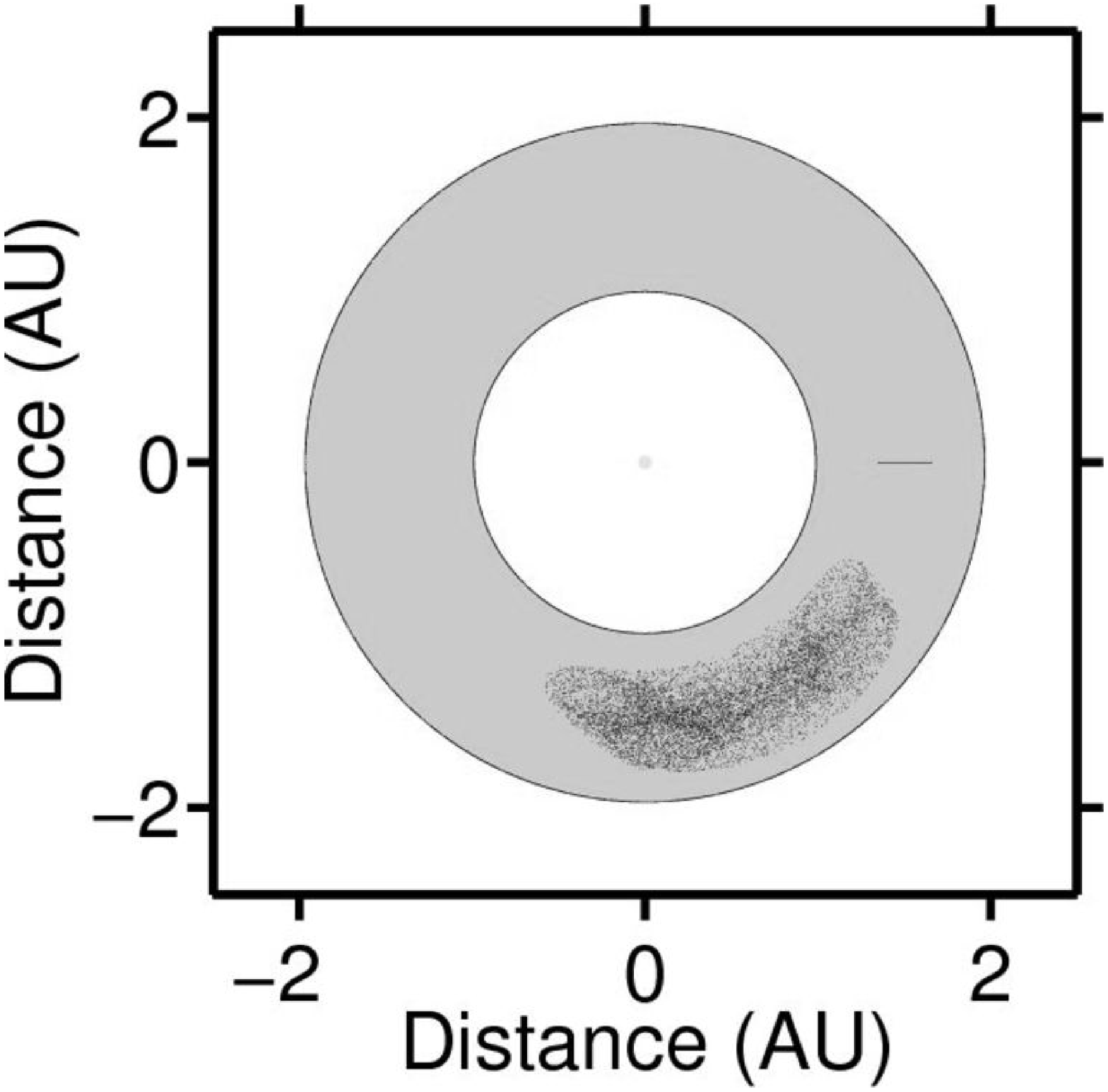,width=0.40\linewidth} &
\epsfig{file=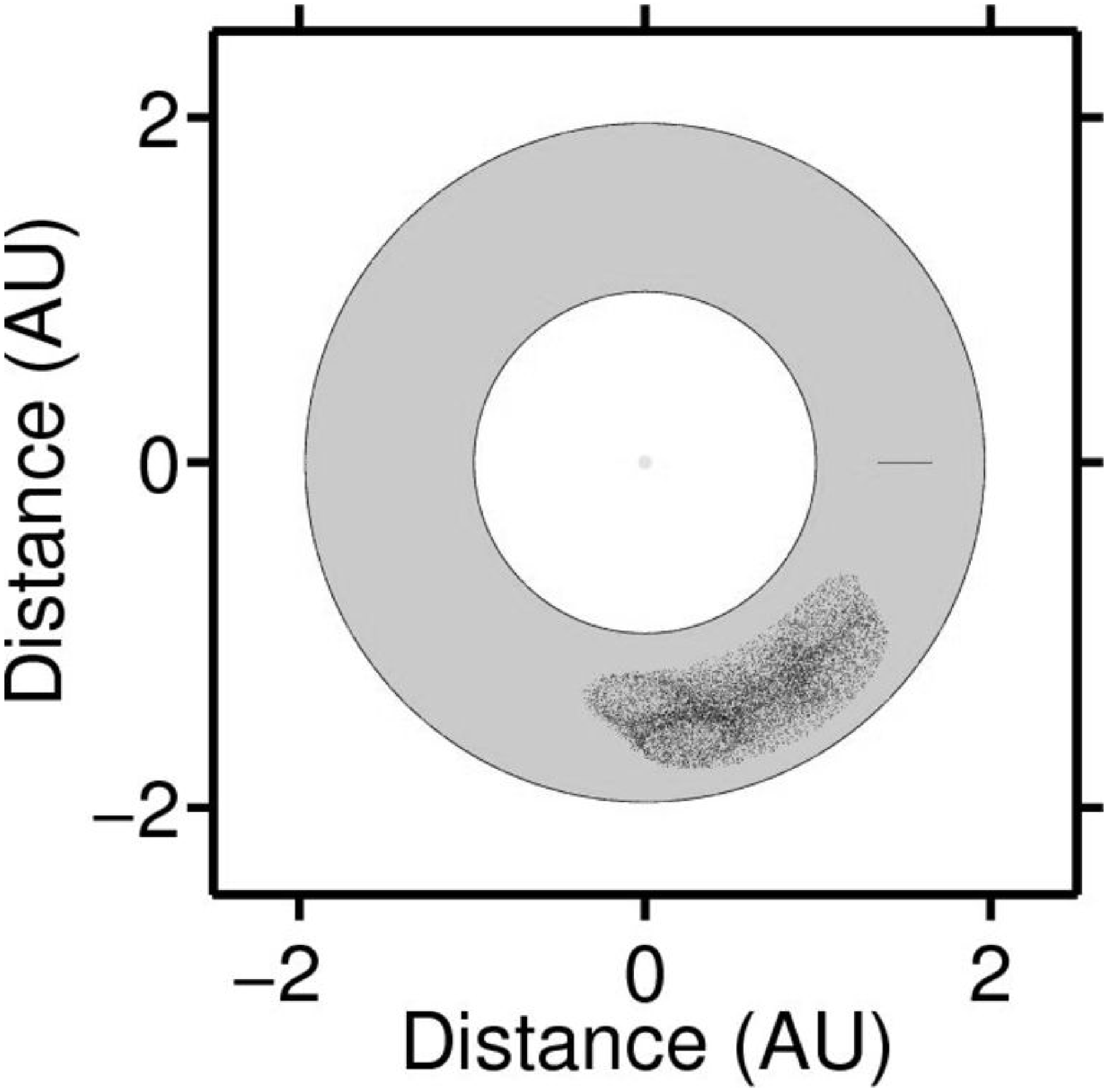,width=0.40\linewidth}
\end{tabular}
\caption{
Orbital stability simulations with HD~23079~b initially placed at apastron
with $a_p$ = 1.503~AU, $e_p$ = 0.102 and the Earth-mass planets placed at
four different starting angles, which are:
45$^\circ$ (top left), 90$^\circ$ (top right), 270$^\circ$ (bottom left),
and 315$^\circ$ (bottom right).  Using a rotating coordinate system,
HD~23079~b moves along the thin line.  The Earth-mass Trojan planets,
which give rise to the ``banana-shaped" area at L4 or L5, remain within
the HZ for at least 10$^6$ years.
}
\end{figure*}

%%%%%%%%%%%%%%%%%%%%%%%%%%%%%%%%%%%%%%%%%%%%%%%%%%%%%%%%%%%%%%%%%

\clearpage

%%% *** Fig.5
%%%%%%%%%%%%%%%%%%%%%%%%%%%%%%%%%%%%%%%%%%%%%%%%%%%%%%%%%%%%%%%%%
\begin{figure*}
\centering
\begin{tabular}{c}
\epsfig{file=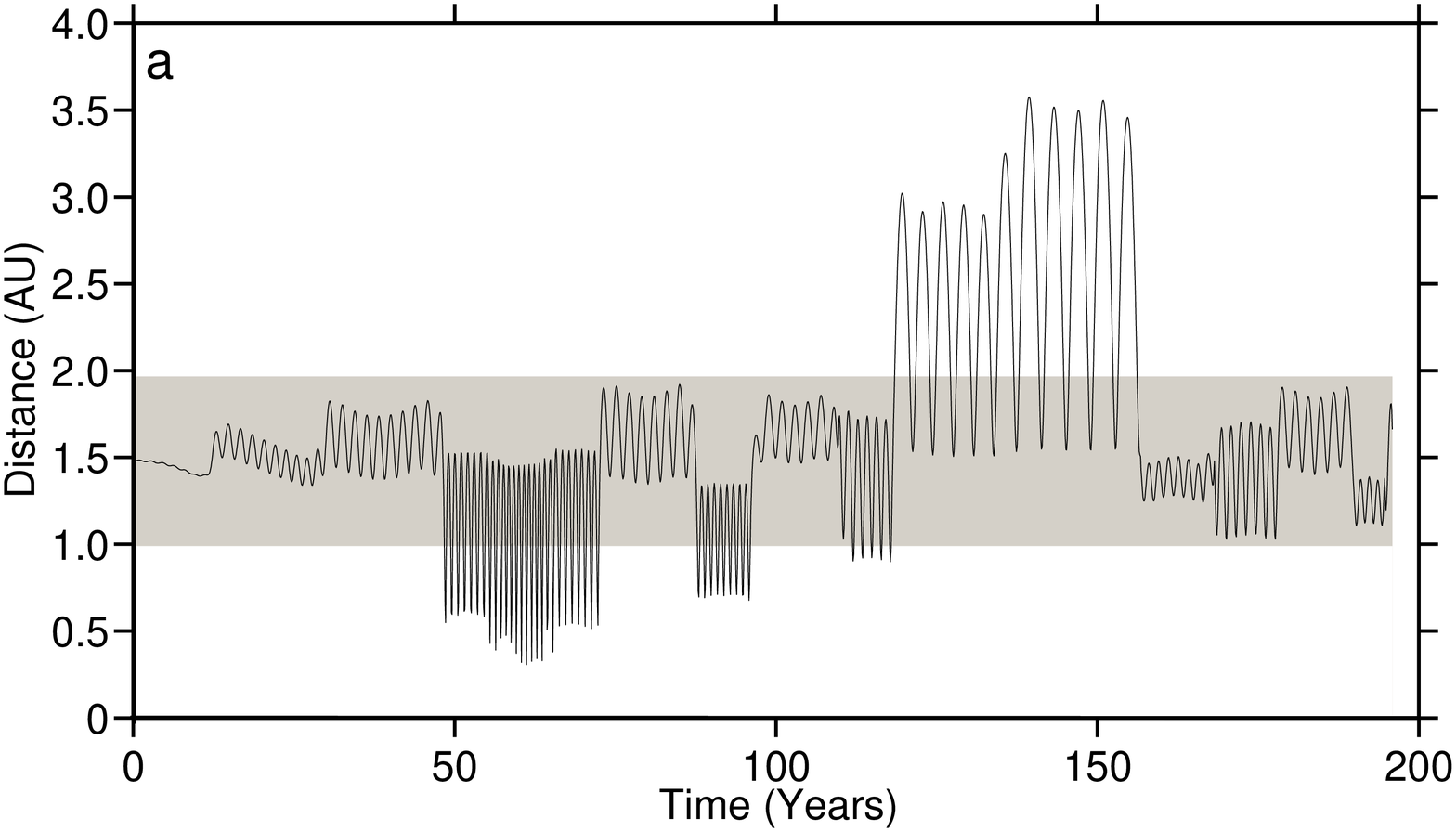,width=0.45\linewidth} \\
\epsfig{file=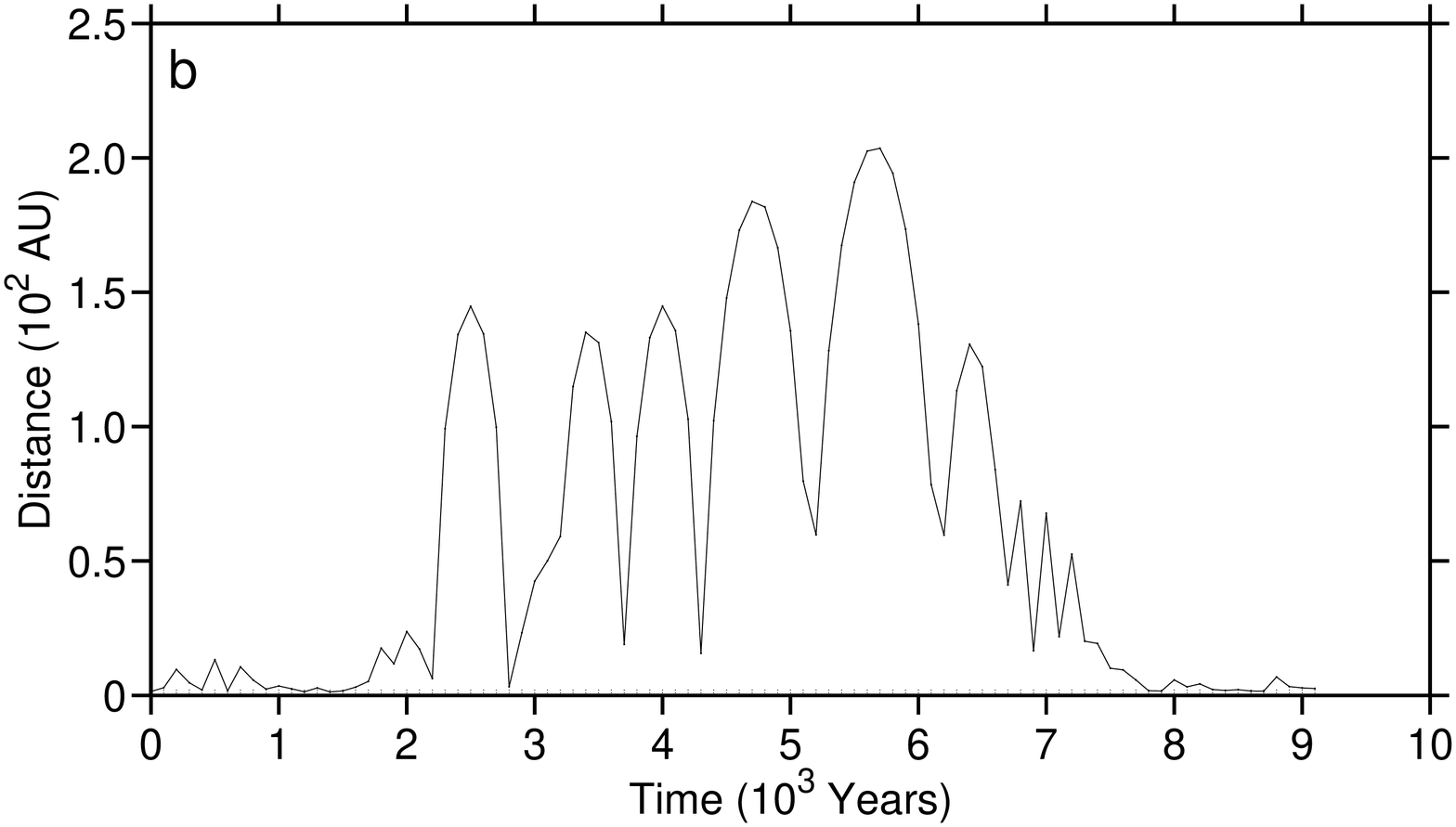,width=0.45\linewidth} \\
\epsfig{file=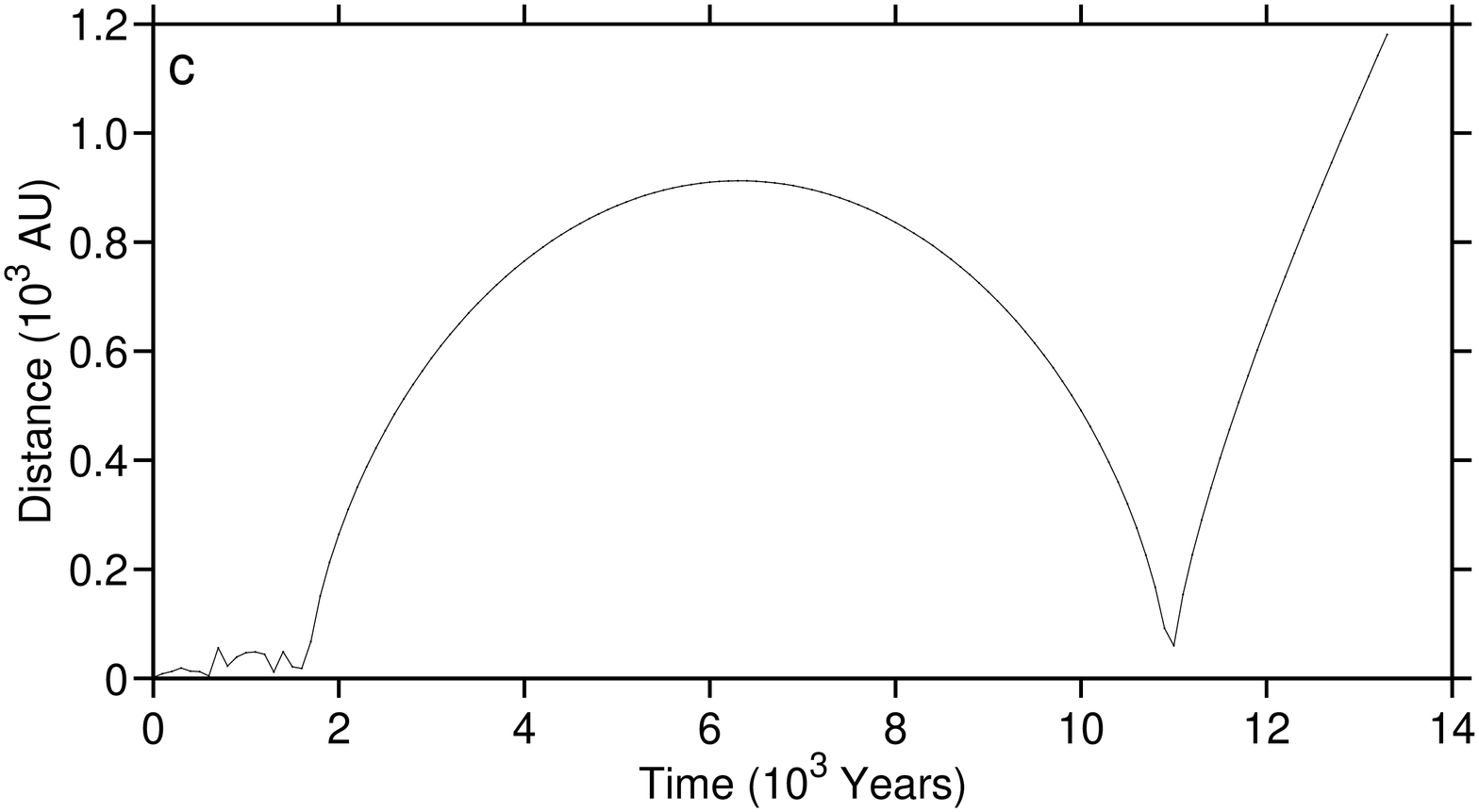,width=0.45\linewidth} \\
\epsfig{file=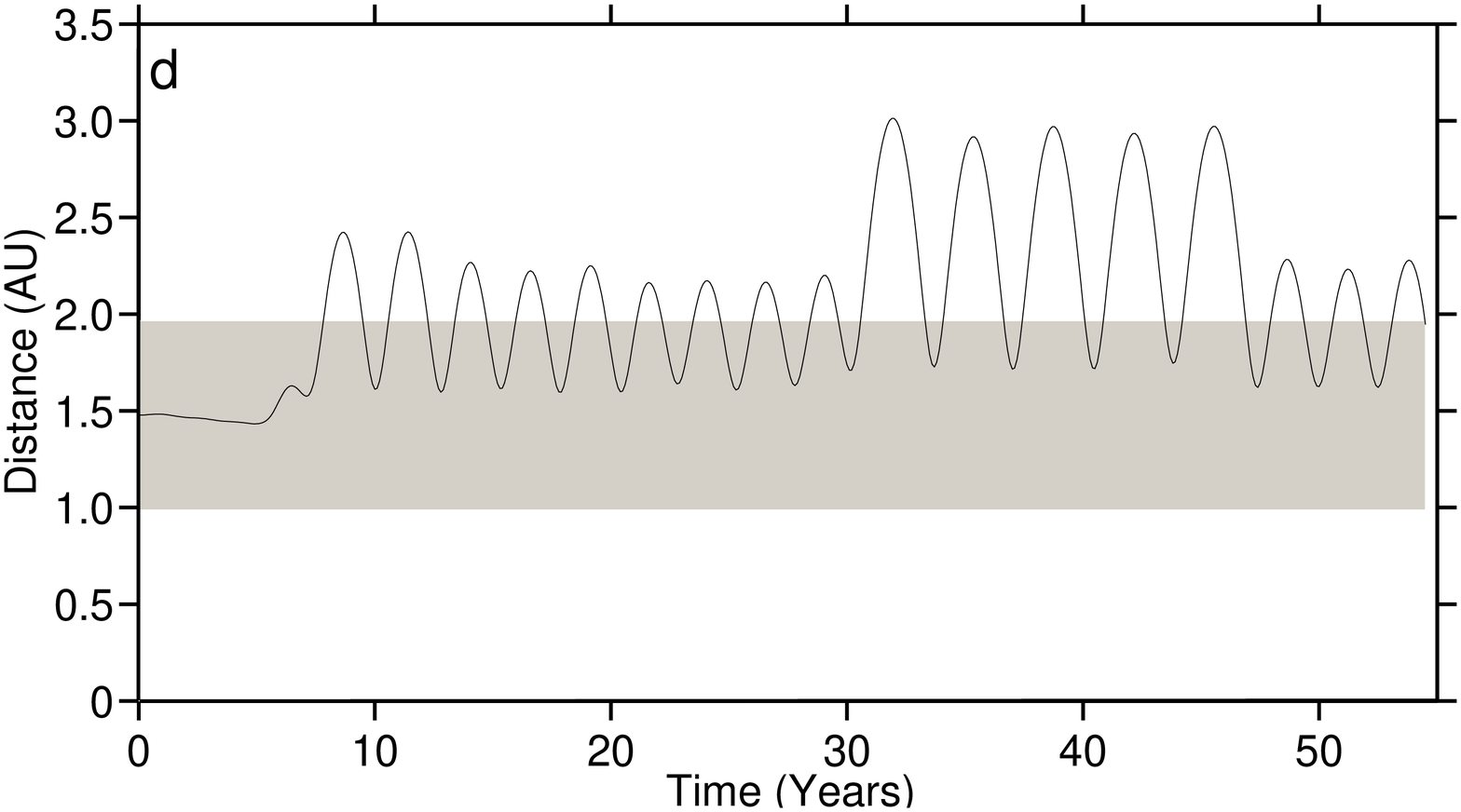,width=0.45\linewidth} \\
\epsfig{file=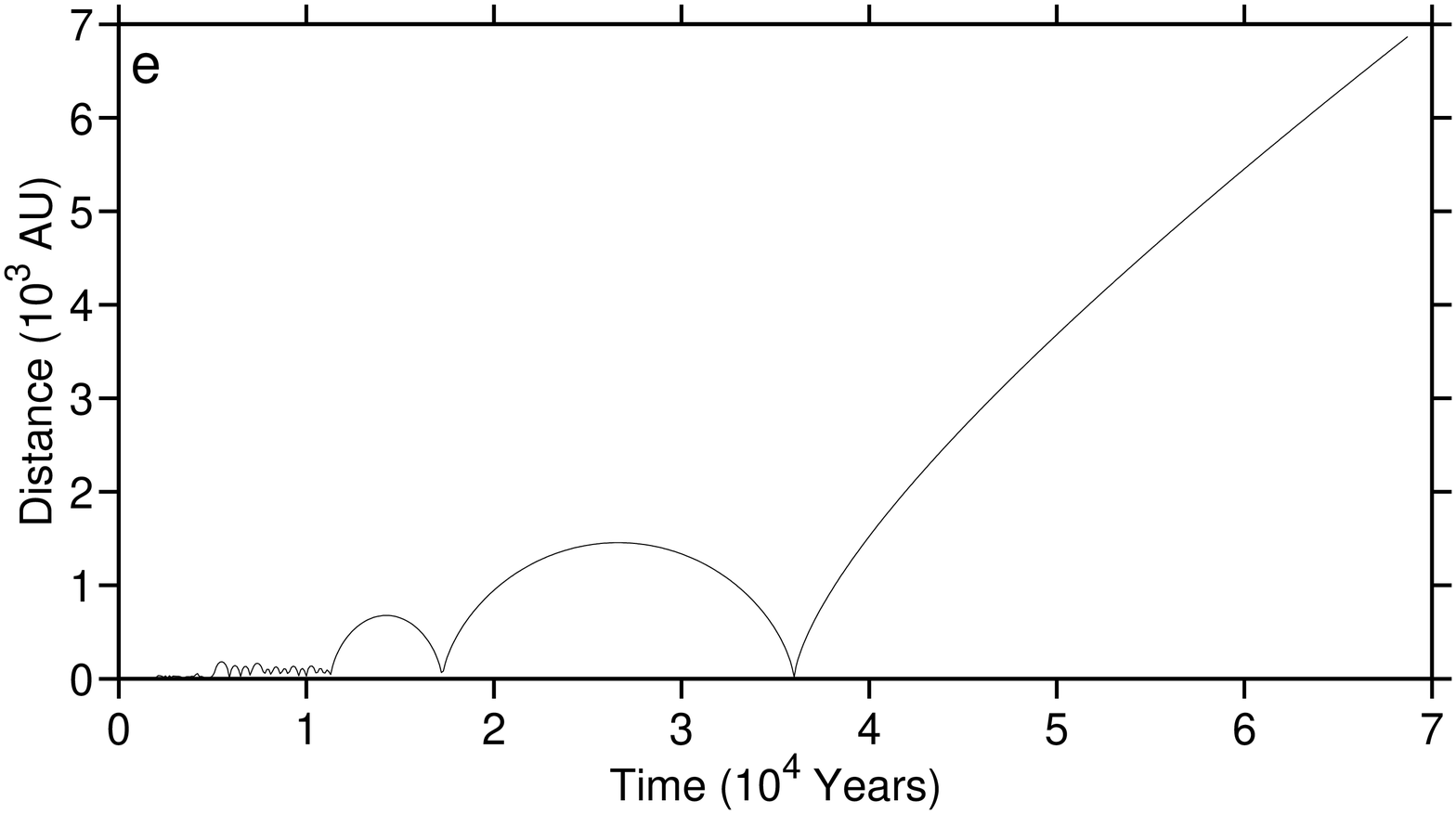,width=0.45\linewidth}
\end{tabular}
\caption{
Orbital stability simulations with HD~23079~b initially placed at
periastron for Earth-mass planets placed at starting angles of
180$^\circ$.  The model simulations differ regarding the selected
values for the orbital parameters $a_p$ and $e_p$ of the giant
planet HD~23079~b (see also Table 2 for further information).
The respective value pairs ($a_p$, $e_p$) are: (1.503, 0.102),
(1.596, 0.071), (1.596, 0.102), (1.596, 0.133), and (1.689, 0.102)
for panel $a$, $b$, $c$, $d$, and $e$, respectively, with $a_p$
in AU.  The grey domains (only visible in panel $a$ and $d$)
depict HD~23079's stellar HZ; see Table~2 for the times of
first exit of the planet from the HZ.  Note the vast differences
in the extent of the $x$ and $y$-axes.
}
\end{figure*}

%%%%%%%%%%%%%%%%%%%%%%%%%%%%%%%%%%%%%%%%%%%%%%%%%%%%%%%%%%%%%%%%%

\clearpage

%%% *** Fig.6
%%%%%%%%%%%%%%%%%%%%%%%%%%%%%%%%%%%%%%%%%%%%%%%%%%%%%%%%%%%%%%%%%
\begin{figure*}
\centering
\begin{tabular}{c}
\epsfig{file=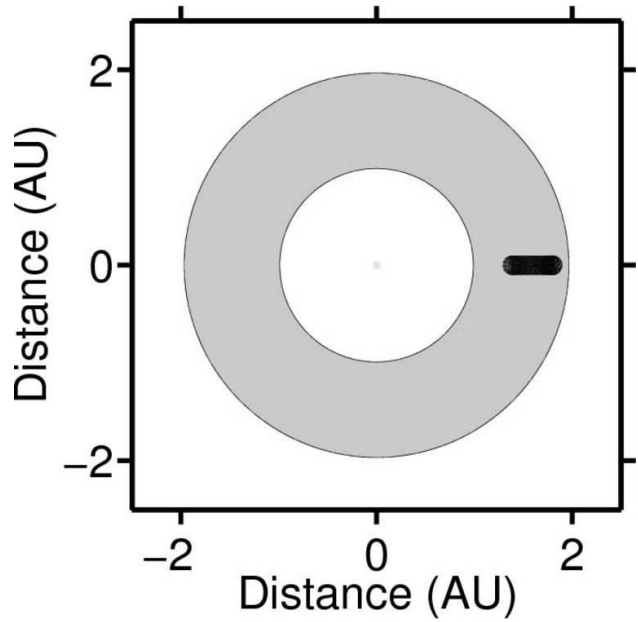,width=0.50\linewidth}
\end{tabular}
\caption{
Orbital stability simulations with HD~23079~b initially placed at periastron
with $a_p$ = 1.596~AU, $e_p$ = 0.133 and the Earth-mass planets placed at
a starting angle of 0$^\circ$.  The Earth-mass planet remains within the HZ
for at least 10$^6$ years.  However, during that time it was captured by
the giant planet and thus became a natural satellite (moon) of that planet,
resulting in the small black area.  Also note the absence of the
``banana-shaped" area at L4 or L5.
}
\end{figure*}

%%%%%%%%%%%%%%%%%%%%%%%%%%%%%%%%%%%%%%%%%%%%%%%%%%%%%%%%%%%%%%%%%

\clearpage

%%% *** Fig.7
%%%%%%%%%%%%%%%%%%%%%%%%%%%%%%%%%%%%%%%%%%%%%%%%%%%%%%%%%%%%%%%%%
\begin{figure*}
\centering
\begin{tabular}{c}
\epsfig{file=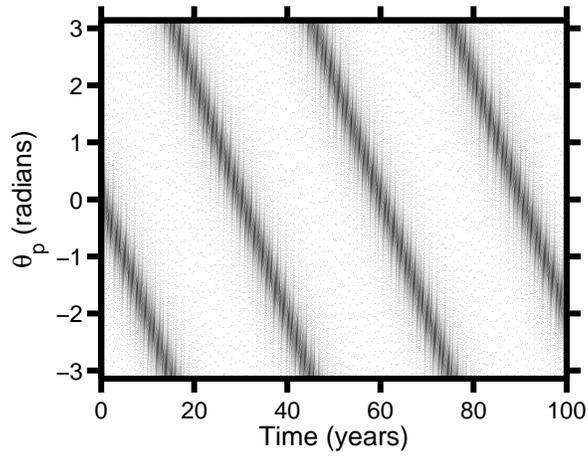,width=0.50\linewidth}
\end{tabular}
\caption{
Using the giant planet as reference and origin of the coordinate system,
we measure the angle $\theta_p$ of the captured terrestrial planet,
which thus became a moon, in a sidereal frame (with $\theta_p = 0$
corresponding to the 3~o'clock position).  With a uniform data
sampling rate, the moon will be more likely recorded at or near apogee.
Since the orbit of the moon is highly eccentric, there are many more
points when the moon is near apogee compared to when it is near perigee.
}
\end{figure*}

%%%%%%%%%%%%%%%%%%%%%%%%%%%%%%%%%%%%%%%%%%%%%%%%%%%%%%%%%%%%%%%%%

\clearpage

%%% *** Fig.8
%%%%%%%%%%%%%%%%%%%%%%%%%%%%%%%%%%%%%%%%%%%%%%%%%%%%%%%%%%%%%%%%%
\begin{figure*}
\centering
\begin{tabular}{c}
\epsfig{file=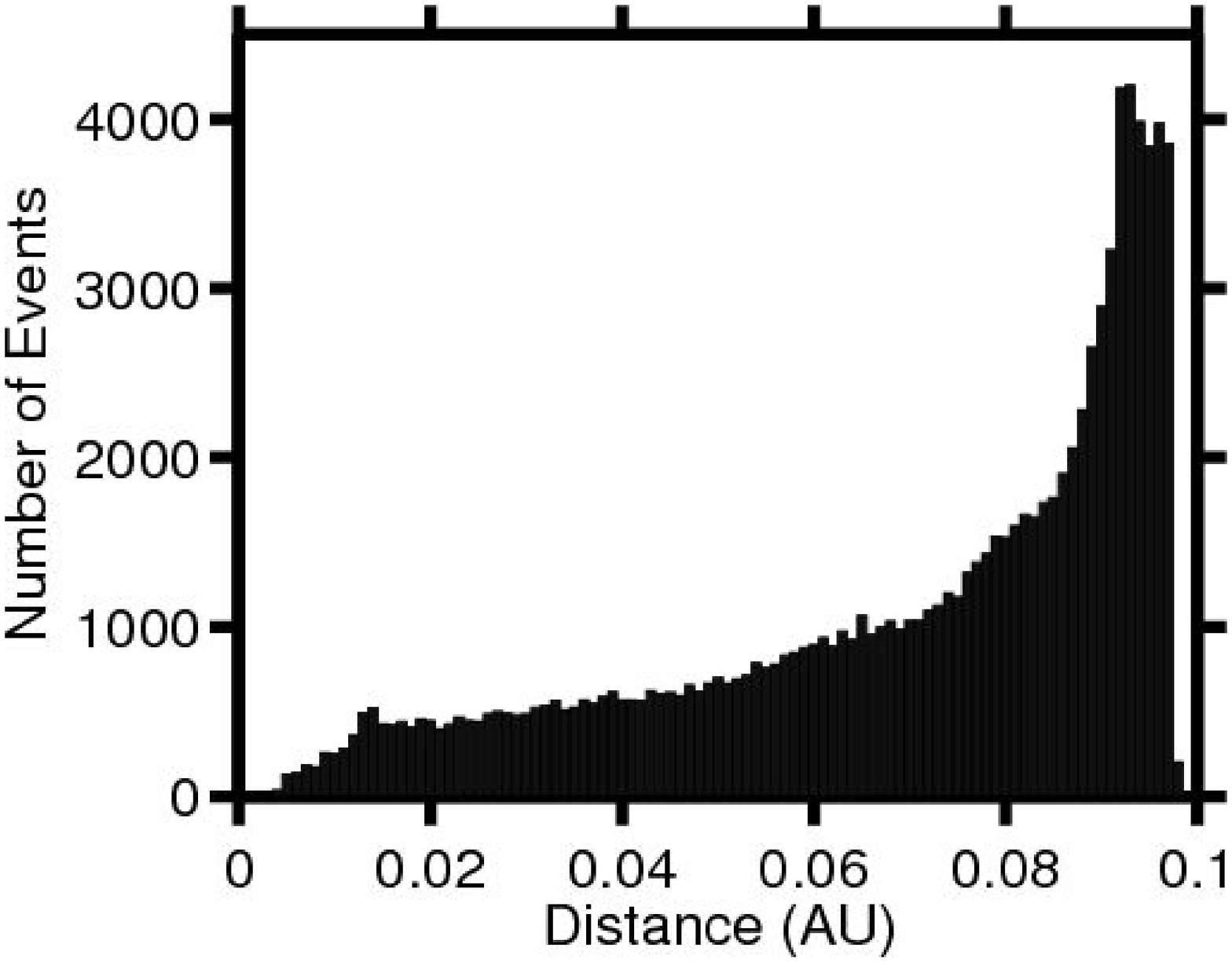,width=0.50\linewidth} \\
\epsfig{file=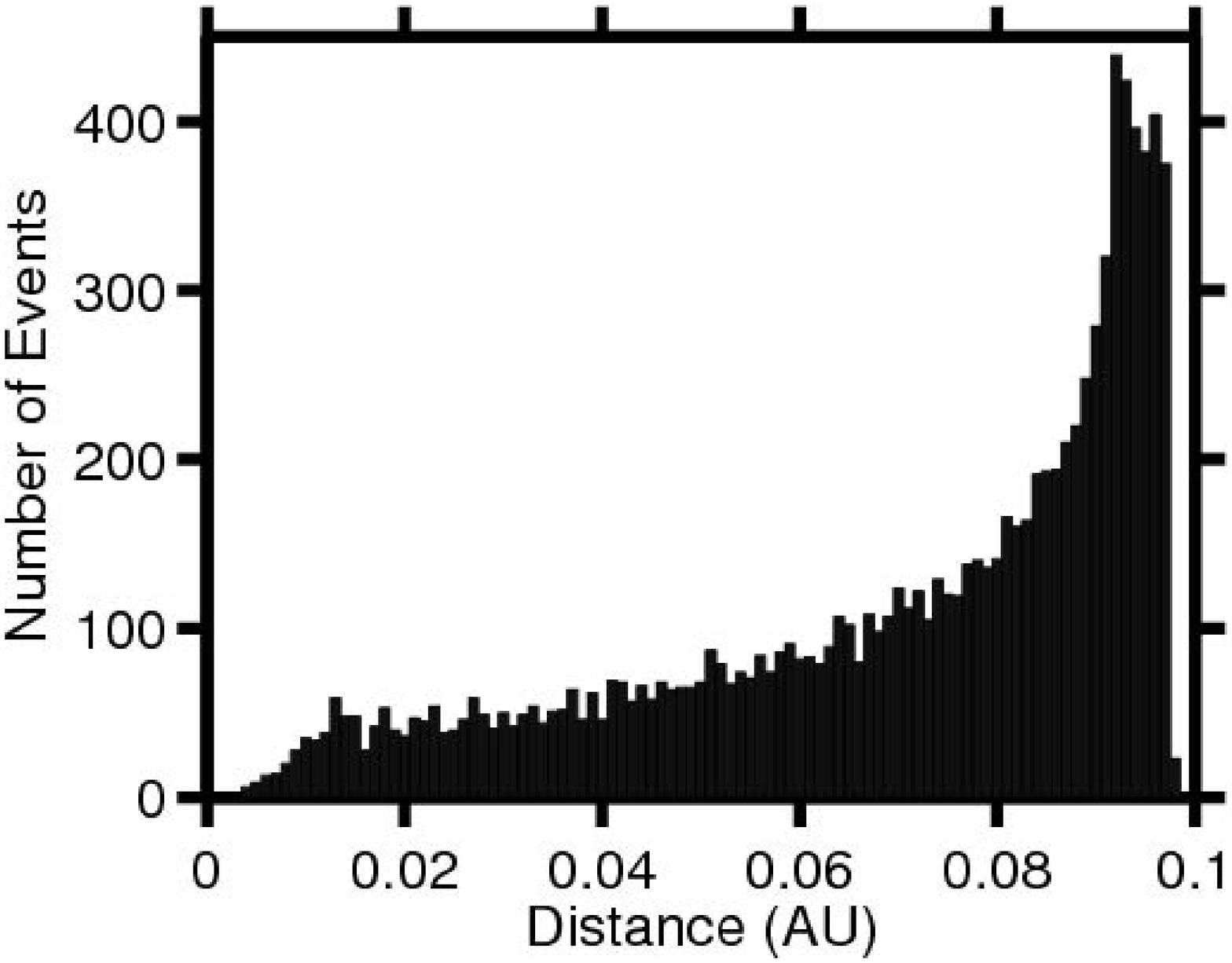,width=0.50\linewidth}
\end{tabular}
\caption{
Histogram displaying the time-dependent distance between the moon and
the giant planet.  Top:  Simulation with an elapsed time of 1000 years
and a data sampling period of 0.01 years.  Bottom:  Simulation with an
elapsed time of $1 \times 10^6$ years and a data sampling period of
100 years.  The comparison between both figures shows that there
is little, if any, evolution of the system over the total time of
simulation.
}
\end{figure*}

%%%%%%%%%%%%%%%%%%%%%%%%%%%%%%%%%%%%%%%%%%%%%%%%%%%%%%%%%%%%%%%%%

\clearpage

%%%%%%%%%%%%%%%%%%%%%%%%%%%%%%%%%%%%%%%%%%%%%%%%%%%%%%%%%%%%%%%%%
%%%%%%%%%%%%%%%%%%%%%%%%%%%%%%%%%%%%%%%%%%%%%%%%%%%%%%%%%%%%%%%%%

\begin{deluxetable}{lcc}
%\tabletypesize{\scriptsize}
\tablecaption{Stellar and Planetary Parameters\label{tab1}}
\tablewidth{0pt}
\tablehead{
Parameter               &  Value                 & Reference
}
\startdata
 Spectral Type          &    F9.5~V              & \cite{gra06}   \\              
 RA                     &       3 39 43.0952     & \cite{esa97}$^{a,b}$ \\      
 DEC                    &  $-$ 52 54 57.017      & \cite{esa97}$^{a,b}$ \\      
 $T_{\rm eff}$~(K)      &  6030  $\pm$ 52        & \cite{rib03}   \\            
 $R$~($R_\odot$)        &  1.106 $\pm$ 0.022     & \cite{rib03}   \\            
 $M$~($M_\odot$)        &  1.10  $\pm$ 0.15      &    $^c$        \\              
 $M_V$                  &  4.42  $\pm$ 0.05      & \cite{esa97}$^{a,b,d}$ \\              
 $M_{\rm bol}$          &  4.25  $\pm$ 0.05      & \cite{esa97}$^{a,b,d}$ \\              
 Distance~(pc)          & 34.60  $\pm$ 0.67      & \cite{esa97}$^{a,b,d}$ \\           
 $M_p {\sin}i$~($M_J$)  &  2.45  $\pm$ 0.21      & \cite{but06}   \\                 
 $P$~(days)             & 730.6  $\pm$ 5.7       & \cite{but06}   \\              
 $a_p$~(AU)             &  1.596 $\pm$ 0.093     & \cite{but06}   \\              
 $e_p$                  &  0.102 $\pm$ 0.031     & \cite{but06}   \\              
\enddata

\tablecomments{
$^a$data from SIMBAD, see        \\
\phantom{XXXXXXXX}~{\tt http://simbad.u-strasbg.fr} \\
\phantom{XXXXXXX}~$^b$adopted from the {\it Hipparcos} catalogue \\
\phantom{XXXXXXX}~$^c$based on spectral type \\
\phantom{XXXXXXX}~$^d$based on parallax $28.90 \pm 0.56$~mas
}
\end{deluxetable}

%%%%%%%%%%%%%%%%%%%%%%%%%%%%%%%%%%%%%%%%%%%%%%%%%%%%%%%%%%%%%%%%%

\clearpage

\begin{deluxetable}{cccccccccc}
\tabletypesize{\scriptsize}
\tablecaption{Time of First Exit (or Ejection) of the Earth-Mass Planet from the HZ (in years)\label{tab2}}
\tablewidth{0pt}
\tablehead{
$a_p$ & $e_p$ & $0^\circ$ & $45^\circ$ & $90^\circ$ & $135^\circ$ & $180^\circ$ & $225^\circ$ & $270^\circ$ & $315^\circ$ \\
(AU)  & ...   & ...       & ...        & ... & ... & ... & ... & ... & ...
}
\startdata
1.503 & 0.071 & 3.26E$-01$~U & ...          & ...          & 1.67E$+02$~L & 1.05E$+02$~L & 5.38E$+01$~L & ...          & ...          \\
1.503 & 0.102 & 1.84E$+00$~C & 2.30E$+05$~U & 7.09E$+05$~U & 8.51E$+01$~U & 4.84E$+01$~L & 4.27E$+01$~U & 8.78E$+05$~U & 2.59E$+05$~U \\
1.503 & 0.133 & 1.07E$+00$~L & 3.46E$+02$~L & 6.13E$+02$~L & 9.72E$+01$~U & 6.04E$+01$~U & 4.30E$+01$~U & 4.78E$+02$~U & 1.85E$+03$~U \\
1.596 & 0.071 & 1.30E$-03$~C & 2.33E$+02$~U & 1.54E$+01$~U & 9.48E$+00$~U & 7.93E$+00$~U & 5.59E$+00$~U & 3.89E$+00$~U & 1.95E$+00$~U \\
1.596 & 0.102 & 4.83E$-01$~U & 1.53E$+01$~U & 1.53E$+01$~U & 1.23E$+01$~U & 7.80E$+00$~U & 5.56E$+00$~U & 3.73E$+00$~U & 1.77E$+00$~U \\
1.596 & 0.133 & ...          & 3.38E$+01$~L & 1.52E$+01$~U & 9.48E$+00$~L & 7.82E$+00$~U & 6.02E$+00$~U & 3.90E$+00$~U & 1.42E$+00$~C \\
1.689 & 0.071 & 1.09E$-01$~C & 8.47E$+00$~U & 8.99E$+00$~U & 7.25E$+00$~L & 4.86E$+00$~U & 4.05E$+00$~U & 2.64E$+00$~U & 3.90E$+01$~L \\
1.689 & 0.102 & 2.85E$-02$~C & 8.02E$+00$~C & 8.56E$+00$~L & 6.73E$+00$~U & 4.78E$+00$~U & 4.38E$+00$~L & 2.45E$+00$~U & 2.18E$+00$~L \\
1.689 & 0.133 & 5.90E$-03$~C & 8.93E$+00$~U & 1.91E$+01$~U & 6.55E$+00$~L & 4.59E$+00$~U & 8.97E$+00$~U & 2.23E$+00$~U & 8.13E$+00$~U \\
\enddata

\tablecomments{
The total time of simulation is 10$^6$ yrs.  The Jupiter-type planet started at the periastron position and the initial velocity
of the Earth-mass planet was computed to begin a circular motion about the star.
U means that the Earth-mass planet crosses the upper limit of the HZ given as 1.9662~AU;
whereas L means that the Earth-mass planet crosses the lower limit of the HZ given as 0.9896~AU.  C means that the Earth-mass planet
has a close encounter with the giant planet, possibly resulting in a collision; therefore, the simulation was discontinued.  If no data
are given, the simulation lasted beyond 10$^6$ yrs without exiting the HZ.}

\end{deluxetable}

%%%%%%%%%%%%%%%%%%%%%%%%%%%%%%%%%%%%%%%%%%%%%%%%%%%%%%%%%%%%%%%%%

\clearpage

\begin{deluxetable}{cccccccccc}
\tabletypesize{\scriptsize}
\tablecaption{Time of First Exit (or Ejection) of the Earth-Mass Planet from the HZ (in years)\label{tab3}}
\tablewidth{0pt}
\tablehead{
$a_p$ & $e_p$ & $0^\circ$ & $45^\circ$ & $90^\circ$ & $135^\circ$ & $180^\circ$ & $225^\circ$ & $270^\circ$ & $315^\circ$ \\
(AU)  & ...   & ...       & ...        & ... & ... & ... & ... & ... & ...
}
\startdata
1.503 & 0.071 & 8.50E$-01$~U & ...          & ...          & 2.95E$+01$~U & 5.08E$+01$~L & 4.01E$+01$~U & ...          & ...          \\
1.503 & 0.102 & 1.45E$+00$~U & ...          & ...          & 9.94E$+01$~L & 2.32E$+01$~L & 2.11E$+01$~U & ..           & ...          \\
1.503 & 0.133 & 9.33E$-01$~L & 1.28E$+03$~L & 1.46E$+03$~U & 2.54E$+01$~L & 4.99E$+01$~U & 8.52E$+00$~U & 4.19E$+03$~U & 7.58E$+02$~U \\
1.596 & 0.071 & 7.07E$+00$~L & 2.57E$+02$~U & 3.27E$+01$~L & 5.60E$+01$~C & 6.61E$+00$~U & 4.97E$+00$~U & 2.86E$+01$~U & 4.18E$+01$~U \\
1.596 & 0.102 & 1.33E$+01$~U & 3.44E$+01$~U & 5.33E$+01$~U & 5.34E$+01$~U & 6.24E$+00$~C & 4.77E$+00$~U & 6.49E$+01$~U & 2.44E$+01$~U \\
1.596 & 0.133 & 7.07E$+00$~U & 2.05E$+01$~U & 7.08E$+01$~U & 3.23E$+01$~U & 1.73E$+01$~U & 5.07E$+00$~U & 2.26E$+01$~U & 2.08E$+01$~U \\
1.689 & 0.071 & 1.89E$+01$~C & 3.61E$+01$~U & 7.76E$+00$~U & 6.45E$+00$~U & 5.17E$+00$~U & 3.61E$+00$~U & 3.61E$+00$~U & 1.38E$+00$~U \\
1.689 & 0.102 & 7.48E$+00$~L & 1.00E$+01$~U & 7.59E$+00$~U & 7.32E$+00$~U & 5.52E$+00$~L & 3.46E$+00$~U & 3.32E$+00$~U & 1.23E$+00$~U \\
1.689 & 0.133 & 1.87E$+01$~U & 9.64E$+00$~U & 7.41E$+00$~U & 7.02E$+00$~L & 5.53E$+01$~L & 3.27E$+00$~U & 3.44E$+00$~L & 1.08E$+00$~U \\
\enddata

\tablecomments{
The total time of simulation is 10$^6$ yrs.  The Jupiter-type planet started at the apastron position and the initial velocity
of the Earth-mass planet was computed to begin a circular motion about the star.  For further information see notes of Table~2.}

\end{deluxetable}

\end{document}